\documentclass[english,aps,nofootinbib,longbibliography,notitlepage,superscriptaddress,prd,preprintnumbers,showkeys,twocolumn]{revtex4-1}
\usepackage[latin9]{inputenc}
\usepackage{babel}
\usepackage{float}
\usepackage{amsmath}
\usepackage{amssymb}
\usepackage{todonotes}
\usepackage[mathscr]{eucal}
\usepackage{graphicx}
\usepackage{esint}
\usepackage{amsbsy}
 \usepackage{hyperref}
\usepackage{comment}
\usepackage{color}
\usepackage{microtype}
\usepackage{cleveref}
\usepackage{breakurl}
\usepackage{ulem}
\usepackage{bbm}
\usepackage{mathtools}

\makeatletter

\pdfpageheight\paperheight
\pdfpagewidth\paperwidth

\@ifundefined{textcolor}{}{%
 \definecolor{BLACK}{gray}{0}
 \definecolor{WHITE}{gray}{1}
 \definecolor{RED}{rgb}{1,0,0}
 \definecolor{GREEN}{rgb}{0,1,0}
 \definecolor{BLUE}{rgb}{0,0,1}
 \definecolor{CYAN}{cmyk}{1,0,0,0}
 \definecolor{MAGENTA}{cmyk}{0,1,0,0}
 \definecolor{YELLOW}{cmyk}{0,0,1,0}
}

\newcommand{\dd}{\mathrm{d}}

\newcommand{\gsimm}{\raise.3ex\hbox{$>$\kern-.75em\lower1ex\hbox{$\sim$}}}
\newcommand{\lsimm}{\raise.3ex\hbox{$<$\kern-.75em\lower1ex\hbox{$\sim$}}}

\newcommand{\be}{\begin{equation}}
\newcommand{\ee}{\end{equation}}
\newcommand{\ba}{\begin{eqnarray}}
\newcommand{\ea}{\end{eqnarray}}

\newcommand{\bea}{\begin{eqnarray*}}
\newcommand{\eea}{\end{eqnarray*}}

\makeatother

\makeatletter
\DeclareRobustCommand{\rcite}[1]{%
  \rcite@aux#1,\@nil{#1}%
}
\def\rcite@aux#1,#2\@nil#3{%
  \if\relax#2\relax
    Ref.~\cite{#3}%
  \else
    Refs.~\cite{#3}%
  \fi
}
\makeatother

\hypersetup{
    colorlinks = true,
    citecolor = {red},
    linkcolor = {blue},
    urlcolor = {blue},
}

\begin{document}

\title{Neutron star merger GW170817 strongly constrains doubly coupled bigravity}

\author{Yashar Akrami}
\email{akrami@lorentz.leidenuniv.nl}
\affiliation{Lorentz Institute for Theoretical Physics, Leiden University, P.O. Box 9506, 2300 RA Leiden, The Netherlands}

\author{Philippe Brax}
\email{philippe.brax@ipht.fr}
\affiliation{Institut de Physique Th{\'e}orique, Universit{\'e} Paris-Saclay, CEA, CNRS, F-91191 Gif/Yvette Cedex, France}

\author{Anne-Christine Davis}
\email{acd@damtp.cam.ac.uk}
\affiliation{DAMTP, Centre for Mathematical Sciences, University of Cambridge, CB3 0WA, United Kingdam}

\author{Valeri Vardanyan}
\email{vardanyan@lorentz.leidenuniv.nl}
\affiliation{Lorentz Institute for Theoretical Physics, Leiden University, P.O. Box 9506, 2300 RA Leiden, The Netherlands}
\affiliation{Leiden Observatory, Leiden University, P.O. Box 9513, 2300 RA Leiden, The Netherlands}

\begin{abstract}
We study the implications of the recent detection of gravitational waves emitted by a pair of merging neutron stars and their electromagnetic counterpart, events GW170817 and GRB170817A, on the viability of the doubly coupled bimetric models of cosmic evolution, where the two metrics couple directly to matter through a composite, effective metric. We demonstrate that the bounds on the speed of gravitational waves place strong constraints on the doubly coupled models, forcing either the two metrics to be proportional at the background level or the models to become singly coupled. Proportional backgrounds are particularly interesting as they provide stable cosmological solutions with phenomenologies equivalent to that of $\Lambda$CDM at the background level as well as for linear perturbations, while nonlinearities are expected to show deviations from the standard model.
\end{abstract}


\maketitle

\tableofcontents

\section{Introduction}\label{sec:intro}

The discovery of the late-time cosmic acceleration~\cite{Riess:1998cb,Perlmutter:1998np} (see Refs.~\cite{Caldwell:2009ix,Weinberg:2012es,Joyce:2014kja,Bull:2015stt} for recent comprehensive reviews on the subject) triggered a wide interest in modifications of general relativity (see, e.g., Refs.~\cite{2010deto.book.....A,Clifton:2011jh} for reviews). Among these modifications to gravity, the bimetric theory of ghost-free, massive gravity is of particular interest. It stands out especially because of the strong theoretical restrictions on the possibilities for constructing a healthy theory of this type. Indeed, historically it has proven to be difficult to invent a healthy theory of massive, spin-2 field beyond the linear regime. The linearized theory has been  known for a long time~\cite{Fierz:1939ix}, while at the fully nonlinear level the theory has been discovered only recently by constructing the ghost-free\footnote{See, however, Ref.~\cite{Konnig:2016idp} for a discussion of the possibility of constructing viable theories of massive gravity in the presence of ghosts.} theory of massive gravity~\cite{deRham:2010ik,deRham:2010kj,Hassan:2011vm,Hassan:2011hr,deRham:2011rn,deRham:2011qq,Hassan:2011tf,Hassan:2011ea,Hassan:2012qv,Hinterbichler:2012cn}. This development has also naturally led to the healthy theory of interacting, spin-2 fields, i.e. the theory of ghost-free, massive bigravity~\cite{Hassan:2011zd}; see Refs.~\cite{deRham:2014zqa,Hinterbichler:2011tt,Schmidt-May:2015vnx,Solomon:2015hja,Hinterbichler:2017sbd} for reviews.

Over the past decade, there has been a substantial effort directed towards understanding the cosmological behavior of bimetric models,\footnote{See Ref.~\cite{Luben:2016lku} for viable background cosmologies of theories with more than two spin-2 fields.} both theoretically and observationally. Particularly, it has been shown that bigravity admits Friedman-Lema\^{i}tre-Robertson-Walker (FLRW) cosmologies\footnote{See Ref.~\cite{Nersisyan:2015oha} and references therein for bimetric cosmologies with other types of background metrics.} which perfectly agree with cosmological observations at the background level~\cite{Volkov:2011an,Comelli:2011zm,vonStrauss:2011mq,Akrami:2012vf,Akrami:2013pna,Konnig:2013gxa,Enander:2014xga,Mortsell:2017fog}. At the level of linear perturbations, the theory has been studied extensively in Refs.~\cite{Comelli:2012db,Khosravi:2012rk,Berg:2012kn,Konnig:2014dna,Solomon:2014dua,Konnig:2014xva,Lagos:2014lca,Cusin:2014psa,Yamashita:2014cra,DeFelice:2014nja,Fasiello:2013woa,Enander:2015vja,Amendola:2015tua,Johnson:2015tfa,Konnig:2015lfa,Lagos:2016gep}, and the cosmological solutions have been shown to suffer from either ghost or gradient instabilities, although the latter can be pushed back to arbitrarily
early times by imposing a hierarchy between the two Planck masses of the theory~\cite{Akrami:2015qga}. It is also conjectured~\cite{Mortsell:2015exa} that the gradient instability might be cured at the nonlinear level due to the presence of the Vainshtein screening mechanism~\cite{Vainshtein:1972sx,Babichev:2013usa} in the theory. The version of the bimetric theory studied in all this work is the so-called {\it singly coupled} scenario, where the matter sector is assumed to couple to only one of the two metrics (spin-2 fields). The metric directly coupled to matter is called the {\it physical metric}, and the other spin-2 field, called the {\it reference metric}, affects the matter sector only indirectly and through its interaction with the physical metric.

In the absence of any theoretical mechanism that forbids the coupling of the matter fields directly to the reference metric, it is natural to go beyond the singly coupled scenarios and study {\it doubly coupled} models, where the two metrics couple to matter either directly or through a composite metric constructed out of the two spin-2 fields. This generalization might look even more natural since the gravity sector of ghost-free bigravity is fully symmetric in terms of the two metrics, and it might feel unnatural to impose the matter sector to break this symmetry by coupling only to one metric.\footnote{Note also that such theories do not necessarily violate the equivalence principle, and if they do, this may not be an issue. For discussions on the violation of the equivalence principle in theories with both metrics minimally coupled to matter, see Refs.~\cite{Akrami:2013ffa,Akrami:2014lja}. For theories with a composite metric coupled to matter the (weak) equivalence principle is not violated, as all particles move along the geodesics of the composite metric.} Theories of doubly coupled massive gravity and bigravity, and, in particular, their cosmologies, have also been extensively studied~\cite{Hassan:2012wr,Akrami:2013ffa,Tamanini:2013xia,Akrami:2014lja,Yamashita:2014fga,deRham:2014naa,deRham:2014naa,Hassan:2014gta,Enander:2014xga,Solomon:2014iwa,Schmidt-May:2014xla,deRham:2014fha,Gumrukcuoglu:2014xba,Heisenberg:2014rka,Gumrukcuoglu:2015nua,Hinterbichler:2015yaa,Heisenberg:2015iqa,Heisenberg:2015wja,Lagos:2015sya,Melville:2015dba,Heisenberg:2016spl,Brax:2016ssf,Brax:2017hxh,Brax:2017pzt}. It has been shown, particularly, that the dangerous Boulware-Deser (BD) ghost~\cite{Boulware:1973my} reemerges almost always if the same matter fields couple to both metrics. One interesting exception has been proposed in Ref.~\cite{deRham:2014naa}, where an acceptable doubly coupled theory of bimetric gravity has been constructed with matter coupled to a composite metric of the form
\be\label{eq:effmetric}
g_{\mu\nu}^{\mathrm{eff}}=\alpha^{2}g_{\mu\nu}+2\alpha\beta g_{\mu\gamma}(\sqrt{g^{-1}f})^{\gamma}{}_{\nu}+\beta^{2}f_{\mu\nu}\,,
\ee
with $g_{\mu\nu}$ and $f_{\mu\nu}$ being the two metrics of the theory, and $\alpha$ and $\beta$ being two arbitrary constants. Clearly, setting $\beta$ to $0$ ($\alpha$ to $0$) turns the doubly coupled theory into a singly coupled one with $g_{\mu\nu}$ ($f_{\mu\nu}$) being the physical metric. Even though in this case the BD ghost is not completely removed from the theory, it is effective only at high energies above the cutoff scale of the theory,\footnote{This cutoff scale for massive gravity, corresponding to the strong-coupling scale, is $\Lambda_3\equiv(m^2 M_{\rm Pl})^{1/3}$, where $m$ is the graviton mass and $M_\text{Pl}$ is the Planck mass. The cutoff scale can be higher for bigravity~\cite{Akrami:2015qga}.} making it a valid effective field theory at low energies.

This doubly coupled theory has been shown to provide viable and interesting cosmological solutions at the background level~\cite{Enander:2014xga,Lagos:2015sya}, with linear perturbations that are stable at least around specific cosmological backgrounds~\cite{Comelli:2015pua} (see also Refs.~\cite{Gumrukcuoglu:2015nua,Brax:2016ssf,Brax:2017hxh,Brax:2017pzt}). In particular, in contrast to the singly coupled theory, this double coupling admits combinations of proportional metrics at the background level, and interestingly, the effective metric always corresponds to the massless fluctuations around such backgrounds, i.e. it satisfies the linearized Einstein equations. It can further be considered as a nonlinear massless spin-2 field~\cite{Schmidt-May:2014xla}. This means that around proportional backgrounds the theory is equivalent to general relativity at the background level as well as for linear perturbations, and differences from general relativity are expected only at the nonlinear level, at least in the sector coupled to matter. The immediate implication of this feature is that doubly coupled bigravity admits viable and stable cosmologies at least for proportional metrics, which are potentially distinguishable from standard cosmology in the nonlinear regime.\footnote{The linear cosmological perturbations for doubly coupled bigravity around proportional, FLRW backgrounds separate into two decoupled sectors. The first (visible) sector coupled to matter is equivalent to general relativity. The second (hidden) sector is decoupled from matter and is not free from some instabilities. The most dangerous one \cite{Comelli:2015pua,Brax:2016ssf} occurs for vectors, which have a gradient instability in the radiation era. This may jeopardize the perturbativity of the models very early on in the Universe. On the other hand, however, the doubly coupled models with a mass $m\sim H_0$ are expected to have an ultraviolet (UV) cutoff scale of order $\Lambda_3= (H_0^2 M_{\rm Pl})^{1/3}$, which is low
and prevents any reliable description of the physics of bigravity when the horizon scale becomes smaller than $\Lambda^{-1}_3$. Strictly speaking, for bimetric theories $\Lambda_3$ is the cutoff scale in the decoupling limit, and the cutoff scale for the full theory can be higher, contrary to massive gravity. However, since the decoupling limit is not well defined above $\Lambda_3$, we expect the entire theory to need modifications. The $\Lambda_3$ scale happens at a redshift of order $10^{12}$ which is just before big bang nucleosynthesis. The unknown UV completion of doubly coupled bigravity would certainly affect the early-Universe instability. In the late Universe as we consider here, no instability is present and the
decoupled sector can be safely ignored for proportional backgrounds.}  As we show in this paper, proportional metrics are extremely interesting also from the point of view of gravitational waves (GWs), as they are the only cases that survive after the recent measurements of the speed of gravity in addition to the singly coupled models. This provides us with a unique class of bimetric models that are healthy and compatible with all cosmological observations as well as  gravitational wave constraints.

Given the large number of possible modifications to gravity, it is natural to ask how all these theories can be tested and potentially falsified. Several high-precision large-scale structure surveys are planned to come into operation in the very near future, and therefore most attempts so far have focused on studying the cosmological implications of such theories in a hope that the future cosmological surveys will be sufficiently sensitive to judge against or for many of these theories. Notably, however, the recent detection of the GWs originating from a pair of merging neutron stars and the simultaneous detection of their electromagnetic counterpart, events GW170817~\cite{TheLIGOScientific:2017qsa} and GRB 170817A~\cite{Goldstein:2017mmi}, have proven to be able to provide us with an immense amount of knowledge about the landscape of the possible theories of gravity (mainly) through the strong bounds that they have placed on the speed of GWs~\cite{Creminelli:2017sry,Sakstein:2017xjx,Ezquiaga:2017ekz,Baker:2017hug,Nojiri:2017hai,Boran:2017rdn,Amendola:2017orw,Crisostomi:2017lbg,Langlois:2017dyl,Gumrukcuoglu:2017ijh,Heisenberg:2017qka,Kreisch:2017uet,Dima:2017pwp,Peirone:2017ywi,Crisostomi:2017pjs,Linder:2018jil,Kase:2018iwp,Battye:2018ssx} (see also Refs.~\cite{Lombriser:2015sxa,Brax:2015dma,Lombriser:2016yzn,Pogosian:2016pwr,Bettoni:2016mij} for discussions on the consequences of such strong bounds for classes of modified theories of gravity prior to the actual observations).

GWs in bigravity have been studied in Refs.~\cite{DeFelice:2013nba,Saltas:2014dha,Cusin:2014psa,Narikawa:2014fua,Amendola:2015tua,Johnson:2015tfa,Max:2017flc,Brax:2017hxh,Nishizawa:2017nef,Max:2017kdc}, although they have been investigated for the doubly coupled models only in Ref.~\cite{Brax:2017hxh}. In the literature, bigravity models are often considered to be on the safe side with respect to the bounds placed by current observations of GWs. While this holds for singly coupled models, we show in this paper that the bounds on the speed of GWs severely constrain the parameter space of the doubly coupled scenarios. We particularly show that the models which survive the bounds from current gravitational wave observations are the ones for which the two background metrics are proportional, or for the choices of the parameters of the model that render it singly coupled.

We first derive, analytically, the conditions under which bimetric models are safe in terms of the gravitational wave measurements. We then perform a Markov chain Monte Carlo (MCMC) analysis of the parameter space of doubly coupled bigravity by imposing the constraints from geometrical measurements of cosmic history, now taking into account also the constraints from gravitational wave observations. We illustrate that this numerical analysis confirms our analytical arguments.

The paper is organized as follows: In Sec.~\ref{DCB} we summarize the basics of doubly coupled bigravity and its cosmology, and present the equations necessary for studying the background cosmological evolution. Section~\ref{GWspeed} discusses the evolution equations and the speed of GWs in the theory and presents the cosmological conditions that result in the speed equal to the speed of light. Section~\ref{constraints} provides the results of our MCMC scans, and our conclusions are given in Sec.~\ref{conclusions}. Finally, in Appendix~\ref{Tens_modes} we derive the cosmological evolution equations for tensor modes in detail, at the level of the field equations as well as the action.

\section{Cosmology of doubly coupled bigravity}\label{DCB}

The theory of doubly coupled bigravity can be formulated in terms of an action of the form~\cite{deRham:2014naa,Enander:2014xga}
\begin{align}\label{eq:action}
S = &-\frac{M^2_{\text{eff}}}{2}\int{\dd^4x\sqrt{-g}R_g}-\frac{M^2_{\text{eff}}}{2}\int{\dd^4x\sqrt{-f}R_f}\nonumber\\
&+m^2M^2_{\text{eff}}\int{\dd^4x\sqrt{-g}}\sum^{4}_{n = 0}{\beta_n e_n (\sqrt{g^{-1}f})}\nonumber\\
&+S_{\text{matter}}[g^\text{eff}_{\mu\nu},\Psi]\,,
\end{align}
where $g_{\mu\nu}$ and $f_{\mu\nu}$ are the two metrics of the theory with determinants $g$ and $f$, respectively, and standard Einstein-Hilbert kinetic terms. $M_{\text{eff}}$ plays the role of the Planck mass,\footnote{It should be noted that the theory can be formulated in terms of two separate Planck masses $M_g$ and $M_f$ corresponding to the $g$ and $f$ sectors, respectively. As has been shown in Ref.~\cite{Enander:2014xga}, the effective metric in this case does not include any free parameters and has the fixed form $g_{\mu\nu}+2g_{\mu\gamma}(\sqrt{g^{-1}f})^{\gamma}{}_{\nu}+f_{\mu\nu}$. We have chosen the formulation in terms of $M_{\text{eff}}$ with $\alpha$ and $\beta$ being present explicitly since it shows the singly coupled limits of the theory more clearly.} $e_{n}$ are the elementary symmetric polynomials of the matrix $\sqrt{g^{-1}f}$ (see Ref.~\cite{Hassan:2011zd} for their detailed definitions), and the quantities $\beta_n$ $(n=0,...,4)$ are five free parameters determining the strength of the possible interaction terms. The parameter $m$ sets the mass scale of the interactions and is not an independent parameter of the theory as it can be absorbed into the $\beta_{n}$ parameters; $m$ needs to be of the order of $H_{0}$, the present value of the Hubble parameter $H$, in order for the theory to provide self-accelerating solutions consistent with observational data. Matter fields have been shown collectively by $\Psi$, which couple to the effective metric $g^\text{eff}_{\mu\nu}$ defined in Eq. (\ref{eq:effmetric}) in terms of $g_{\mu\nu}$ and $f_{\mu\nu}$ and the two coupling parameters $\alpha$ and $\beta$.

In order to study the cosmological implications of the theory, we assume the background metrics $g_{\mu\nu}$ and $f_{\mu\nu}$ to have the FLRW forms
\begin{align}
\dd s^2_g &= -N^2_g \dd t^2 + a^2_g\dd x_i\dd x^i\,,\label{eq:FLRWg}\\
\dd s^2_f &= -N^2_f \dd t^2 + a^2_f\dd x_i\dd x^i\,,\label{eq:FLRWf}
\end{align}
where $t$ is the cosmic time, $N_g$ and $N_f$ are the lapse functions for $g_{\mu\nu}$ and $f_{\mu\nu}$, respectively, and $a_g$ and $a_f$ are the corresponding scale factors, all functions of $t$ only.

Using the forms (\ref{eq:FLRWg}) and (\ref{eq:FLRWf}) for the background metrics $g_{\mu\nu}$ and $f_{\mu\nu}$, Eq. (\ref{eq:effmetric}) fixes the form of the effective metric $g^\text{eff}_{\mu\nu}$ to
\begin{equation}
\dd s^2_{\text{eff}} = -N^2 \dd t^2 + a^2\dd x_i\dd x^i \,,
\end{equation}
where~\cite{Enander:2014xga}
\begin{align}
N &\equiv \alpha N_g + \beta N_f\, ,\\
a &\equiv \alpha a_g + \beta a_f\, ,
\end{align}
are the lapse and the scale factor of the effective metric, respectively. The dynamics of $g_{\mu\nu}$ and $f_{\mu\nu}$ are governed by their Friedmann equations, which take the forms
\begin{align}
3H^2_g &= \frac{\alpha}{M^2_{\text{eff}}}\rho\frac{a^3}{a^3_g}+H^2_0(\beta_0 + 3\beta_1r+3\beta_2r^2 + \beta_3r^3)\, ,\label{eq:friedg}\\
3H^2_f &= \frac{\beta}{M^2_{\text{eff}}}\rho\frac{a^3}{a^3_f}+H^2_0(\frac{\beta_1}{r^3}+3\frac{\beta_2}{r^2}+3\frac{\beta_3}{r} + \beta_4)\, \label{eq:friedf},
\end{align}
where
\begin{equation}
H_g \equiv \frac{\dot{a_g}}{N_g a_g}\,,~~~~~~~~H_f \equiv \frac{\dot{a_f}}{N_f a_f}\, ,
\end{equation}
are the Hubble parameters for $g_{\mu\nu}$ and $f_{\mu\nu}$, respectively, $\rho$ is the energy density of matter and radiation, the dot denotes a derivative with respect to $t$, and
\begin{equation}
r \equiv \frac{a_{f}}{a_{g}}\,
\end{equation}
is the ratio of the two scale factors $a_{f}$ and $a_{g}$. We have also fixed $m$ to $H_{0}$ in the two Friedmann equations, as we are interested in self-accelerating solutions for which $m\sim H_{0}$.

In addition to the two Friedmann equations (\ref{eq:friedg}) and (\ref{eq:friedf}), the consistency of the theory requires the Bianchi constraint~\cite{Enander:2014xga}
\begin{equation}\label{eq:Bianchiconst}
\frac{N_f}{N_g} = \frac{\dot{a_f}}{\dot{a_g}} \rightarrow H_g = r H_f \,
\end{equation}
to be satisfied.\footnote{Note that the Bianchi constraint gives two branches of solutions. The one we consider here is the so-called dynamical branch. See Refs.~\cite{Enander:2014xga,Lagos:2015sya} for the discussion of the second, algebraic branch.} Having introduced the effective lapse and scale factor $N$ and $a$, one can naturally introduce an effective Hubble parameter associated with the effective metric $g_{\mu\nu}^\text{eff}$,
\begin{equation}
H \equiv \frac{\dot{a}}{N a}\, ,
\end{equation}
which satisfies its own effective Friedmann equation~\cite{Enander:2014xga},
\begin{equation}
H^2 = \frac{\rho}{6M^2_{\text{eff}}}(\alpha + \beta r)(\alpha + \frac{\beta}{r}) + H^2_0\frac{B_0 + r^2 B_1}{6(\alpha + \beta r)^2}\,,\label{eq:friedeff}
\end{equation}
where we have also introduced
\begin{align}
B_0 &\equiv \beta_0 + 3\beta_1 r + 3 \beta_2 r^2 + \beta_3 r^3\,,\\
B_1 &\equiv \frac{\beta_1}{r^3} + 3\frac{\beta_2}{r^2} + 3\frac{\beta_3}{r} + \beta_4 \,.
\end{align}
Equation (\ref{eq:friedeff}) is obtained by adding the two Friedman equations (\ref{eq:friedg}) and (\ref{eq:friedf}), and applying the Bianchi constraint (\ref{eq:Bianchiconst}). The effective Hubble parameter $H$ can be written in terms of $H_g$ or $H_f$ as
\begin{equation}
H=\frac{H_g}{\alpha+\beta r}=\frac{rH_f}{\alpha+\beta r}\,.
\end{equation}

In addition to the Friedmann equation for $H$, by again using the Bianchi constraint (\ref{eq:Bianchiconst}) and now subtracting the two Friedmann equations (\ref{eq:friedg}) and (\ref{eq:friedf}) we arrive at the algebraic condition
\begin{equation}\label{eq:algebraic_constraint}
\frac{\rho}{M^2_{\text{eff}}}(\alpha + \beta r)^3 (\alpha - \frac{\beta}{r})+H^2_0(B_0 - r^2 B_1) = 0 \,.
\end{equation}

The energy-momentum tensor for matter and radiation is covariantly conserved with respect to the effective metric, which means that the energy density $\rho$ satisfies the continuity equation
\begin{equation}
\dot{\rho}+3\frac{\dot{a}}{a}(\rho + p) = 0\,.
\end{equation}
This motivates us to introduce $x\equiv\ln a$, the number of $e$-folds in terms of the effective scale factor $a$, as a time coordinate. In terms of $x$, we can recover the usual behavior of the matter and radiation energy densities
\begin{equation}
\rho_{\text{M}} = \rho^{(0)}_{\text{M}}e^{-3x}\,,~~~~~\rho_{\text{R}} = \rho^{(0)}_{\text{R}}e^{-4x}\,,
\end{equation}
assuming that these two components are conserved separately. Here, $\rho^{(0)}_{\text{M}}$ and $\rho^{(0)}_{\text{R}}$ are the current values of the energy densities of matter and radiation, respectively.

It is easy to show that the coupling parameters $\alpha$ and $\beta$ affect observables only though their ratio $\beta/\alpha$, as we can assume $\alpha\neq0$ without loss of generality\footnote{This is indeed the case because the singly coupled bigravity theories with either of the metrics being coupled to  matter are completely equivalent.} and then rescale $M^2_{\text{eff}}$ by a factor of $1/\alpha^4$. Later in this paper, when discussing the constraints, we use this rescaling freedom and introduce a new parameter
\begin{equation}
\gamma \equiv \frac{\beta}{\alpha} \,,
\end{equation}
which plays the role of the only extra parameter for doubly coupled models compared to the singly coupled ones. Identifying the effective Planck mass $M_\text{eff}$ with the usual Planck mass $M_\text{Pl}$, our doubly coupled bimetric model now possesses six free parameters, $\beta_n$ with $n=0,...,4$, and $\gamma$. For now, however, let us keep both $\alpha$ and $\beta$ explicit as it allows us to  see explicitly the duality properties of the background dynamics equations as well as the equations governing the propagation speed of the GWs.

Before we proceed with our studies of gravitational waves in the next sections, let us emphasize an important property of the cosmological evolution equations that we presented in this section. As can be seen easily at the level of the action, the theory is symmetric under the simultaneous interchanges $g_{\mu\nu}\leftrightarrow f_{\mu\nu}$, $\beta_n \rightarrow \beta_{4-n}$ and $\alpha\leftrightarrow\beta$ (or $\gamma\rightarrow 1/\gamma$) and therefore all the dynamical equations remain unchanged~\cite{Enander:2014xga}. More concretely, let us consider two sets of parameters $\{\beta_0,\beta_1,\beta_2,\beta_3,\beta_4,\alpha,\beta\} = \{v_0,v_1,v_2,v_3,v_4,v_5,v_6\}$ and $\{\beta_0,\beta_1,\beta_2,\beta_3,\beta_4,\alpha,\beta\} = \{v_4,v_3,v_2,v_1,v_0,v_6,v_5\}$, where $v_{0,...,6}$ are some particular values of the parameters. It is easy to show that the solution of Eq.~(\ref{eq:algebraic_constraint}) for $r$ with the first set of parameter values is identical to the solution for the quantity $\tilde{r}\equiv 1/r$ with the second set of parameter values. Now if we rewrite Eq.~(\ref{eq:friedeff}) in terms of $\tilde{r}$ (note that we do not make an actual interchange $r\rightarrow 1/r$, and we only rewrite the equations in terms of $\tilde{r}$) then for the two distinct sets of parameter values given above the two Friedmann equations are precisely the same. This, for example, implies that when scanning the single-parameter submodel with all the $\beta_n$ parameters turned off except $\beta_1$ the space of all the cosmological solutions that we obtain is fully equivalent to the one for the submodel with only $\beta_3$ turned on (given that we leave $\alpha$ and $\beta$, or equivalently $\gamma$, free). This is a useful observation and helps us reduce the number of cases studied in the next sections.

\section{The speed of gravitational waves}\label{GWspeed}

The spectrum of bimetric theories of gravity contains two gravitons, one massive and one massless, with five and two degrees of freedom, respectively. In order to study the properties of gravitational waves one needs to focus only on tensor modes, i.e. the helicity-2 modes of the gravitons. Massless and massive gravitons have two helicity-2 modes each. It is important to note that in general the two metrics of the theory, $g_{\mu\nu}$ and $f_{\mu\nu}$, each contain a combination of massive and massless modes, and therefore the evolution equations for the $g$ and $f$ tensor modes do not represent directly the evolution of the tensor modes for massive and massless modes. Indeed, it is not possible in general to diagonalize the spectrum of spin-2 perturbations into mass eigenstates, and therefore the notion of mass does not make sense around arbitrary backgrounds~\cite{Schmidt-May:2014xla}. One can specifically show~\cite{Schmidt-May:2014xla} that mass eigenstates can be defined only around proportional metrics by computing the spectrum of linear perturbations and comparing their equations with those of linearized general relativity. Proportional metrics are therefore extremely interesting from this point of view, as the notion of spin-2 mass eigenstates does not exist for other types of backgrounds. As we mentioned in Sec.~\ref{sec:intro}, contrary to the theory of singly coupled bigravity, the doubly coupled theory admits proportional backgrounds (both in vacuum and in the presence of matter). It can be shown additionally that the effective metric of the theory, $g_{\mu\nu}^\text{eff}$, corresponds exactly to the massless mode around such backgrounds, while the massive mode is fully decoupled~\cite{Schmidt-May:2014xla}. This immediately implies that the speed of GWs around proportional backgrounds measured by any detectors must be equal to the speed of light since the detectors see only the effective metric. Such solutions are therefore safe regarding the bounds from the GW observations. We show later in this paper that, in addition to the singly coupled corner of the theory, proportional backgrounds are indeed the {\it only} solutions that survive the bounds from GWs.

As detailed in Appendix~\ref{Tens_modes}, the propagation equations for the $g$ and $f$ tensor modes $h_{g}$ and $h_{f}$ around the cosmological backgrounds are
\begin{widetext}
\begin{eqnarray}
{h}_{g+/\times}^{\prime\prime} &+& \left(\frac{N^{\prime}}{N}-\frac{N_g^{\prime}}{N_g}-\frac{a^{\prime}}{a}+3\frac{a_g^{\prime}}{a_g}\right){h}_{g+/\times}^{\prime} - \frac{N_g^2}{N^2}\frac{a^2}{a_g^2}\nabla^2 h_{g+/\times}+ \frac{N_g^2}{N^2}a^2A(h_{f+/\times} - h_{g+/\times}) = 0\,,\label{eq:propeq_g}\\
{h}_{f+/\times}^{\prime\prime} &+& \left(\frac{N^{\prime}}{N}-\frac{N_f^{\prime}}{N_f}-\frac{a^{\prime}}{a}+3\frac{a_f^{\prime}}{a_f}\right){h}_{f+/\times}^{\prime} - \frac{N_f^2}{N^2}\frac{a^2}{a_f^2}\nabla^2 h_{f+/\times}+ \frac{N_f^2}{N^2}a^2B(h_{g+/\times} - h_{f+/\times}) = 0\,.\label{eq:propeq_f}
\end{eqnarray}
\end{widetext}
Here, the prime denotes a derivative with respect to the conformal time corresponding to the effective metric, $\eta_{\text{eff}}$, which is defined through
\begin{equation}
d\eta^2_{\text{eff}} = dt^2N^2/a^2.
\end{equation}
With this time coordinate the background effective metric reads
\begin{equation}
ds^2_{\text{eff}} = a^2(-d\eta^2_{\text{eff}}+dx^2)\,.\label{met_eff_conf}
\end{equation}
First note that we have written the equations in terms of the time coordinate corresponding to the effective metric and not $g_{\mu\nu}$ or $f_{\mu\nu}$, because the effective metric is the one that couples to matter and therefore plays the role of the physical spacetime metric, used for measuring distances and time intervals. In addition, we chose to work with the conformal time because in this coordinate light rays travel as in a Minkowski spacetime, making $\eta_\text{eff}$ a particularly useful time coordinate for identifying the propagation speeds of the gravitational waves.

We can now read off from Eqs. (\ref{eq:propeq_g}) and (\ref{eq:propeq_f}) the propagation speeds $c_g$ and $c_f$ for the gravitational waves $h_{g}$ and $h_{f}$, respectively, as\footnote{Note that since we are interested in bigravity solutions with the interaction scale $m\sim H_0$ in order to explain cosmic acceleration, the effects of the graviton mass on the speed of the gravitational waves are several orders of magnitude smaller than the sensitivity of current GW detectors. We therefore fully ignore the direct contributions from the mass terms to the speed.}
\begin{eqnarray}
c^2_g &=& \frac{N^2_g}{N^2}(\alpha + \beta r)^2\,,\\\label{cg}
c^2_f &=& \frac{N^2_f}{N^2}(\alpha\frac{1}{r} + \beta)^2\,.\label{cf}
\end{eqnarray}
The ratio of the two speeds is a coordinate-independent quantity and is given by
\begin{equation}
\frac{c_f}{c_g}=b\equiv \frac{1}{r}\frac{N_{f}}{N_{g}}=\frac{1}{r}\frac{\dot a_f}{\dot a_g}\,.\label{eq:bdef}
\end{equation}
As we see later, the quantity $b$ plays a crucial role in the rest of the discussions in this paper.

One should note again that in doubly coupled bigravity one measures neither $h_{g}$ nor $h_{f}$ separately. The tensor modes measured by gravitational wave detectors are the ones corresponding to the effective metric $g_{\mu\nu}^{\mathrm{eff}}$. These observable modes can be written in terms of $h^{(g)}_{ij}$ and $h^{(f)}_{ij}$, the tensor modes of the $g$ and $f$ metrics respectively, as
\begin{equation}
\delta g^{(\text{eff})}_{ij} = a\left( \alpha h^{(g)}_{ij}+ \beta h^{(f)}_{ij} \right)\,,
\end{equation}
where
\begin{eqnarray}
h^{(I)}_{11} &=& a_I h_{I+}\,,\\
h^{(I)}_{12} &=& a_I h_{I\times} = h^{(I)}_{21}\,,\\
h^{(I)}_{22} &=& -a_I h_{I+}\,,
\end{eqnarray}
with $I\in\{g,f\}$ (see Appendix~\ref{Tens_modes} for details).

The recent measurements of the GWs from neutron star mergers have imposed incredibly tight constraints on the speed of gravitons. The relative difference between the two speeds must be smaller than $\sim 10^{-15}$, which is practically $0$. Let us therefore assume that the speed of GWs is exactly the same as the speed of light, and study its implications.

The mentioned bound on the speed of GWs tells us that at least one of the quantities $c_g$ and $c_f$ should be unity (note that $c=1$ in our units). The reason for this is that at least one of the $g$ or $f$ graviton modes should have traveled with the speed of light when arriving at the detector. Keeping this in mind let us first assume that
\begin{itemize}
\item we are in a truly doubly coupled regime (i.e. $\alpha\neq 0$ and $\beta \neq 0$)\,,
\item $r$ is a finite and nonzero quantity,
\item $N_f$ and $N_g$ are finite and nonzero.
\end{itemize}
Let us further set $N=1$ and write the two speeds $c_g$ and $c_f$ as
\begin{eqnarray}
c^2_g &=& \frac{(\alpha + \beta r)^2}{(\alpha+br\beta)^2}\,,\\
c^2_f &=& \frac{(\alpha\frac{1}{r} + \beta)^2}{(\alpha\frac{1}{br}+\beta)^2}\,.
\end{eqnarray}
Now it is clear that, first of all, when $b = 1$, both $c_g$ and $c_f$ become unity. Moreover, when either $c_g$ or $c_f$ is unity, we necessarily have $b = 1$. This then tells us very strongly that in the case of finite and nonzero $N_f$, $N_g$ and $r$, and under the assumption of $\alpha\neq 0$ and $\beta \neq 0$, $b=1$ is the necessary and sufficient condition for compatibility with the GW experiments.

Let us now discuss the validity of the assumptions that we made above. From the Friedmann equation (\ref{eq:friedeff}) we see that both infinite and zero values of $r$ lead to singularity in the observable Hubble function $H$ unless either $\alpha$ or $\beta$ is $0$, i.e. the theory is singly coupled. This means that for physical solutions in the doubly coupled regime $r$ is necessarily finite and nonzero. Additionally, if $N_f = 0$ while $N_g$ is finite and nonzero, we see that $c^2_f = 0$ while $c^2_g = (1+\gamma r)^2,$\footnote{Here we have used the expression for the effective lapse function $1 = \alpha N_g + \beta N_f$} which is not equal to unity unless we are in the singly coupled regime of $\beta = 0$. In exactly the same way the case of $N_g = 0$ while simultaneously $N_f$ is finite and nonzero is excluded. In principle, one should also consider the cases with one of the lapse functions $N_{g,f}$ going to infinity while their ratio is fixed.\footnote{Otherwise, obviously, they cannot satisfy the gauge fixing condition $N=1$.} Note however that such cases not only produce unphysical propagation speeds in both $g$ and $f$ sectors, but they also remove the second-order time derivatives in the tensor propagation equations, hence rendering the initial data from the past lost at one particular instant in time (when the divergence happens). Based on these considerations we can conclude that the cases with $b=0$ or $b\rightarrow \infty$ are excluded.

Finally, as it is expected, in the singly coupled case (say, $\beta = 0$ and $\alpha = 1$), we have $N_g = 1$ and $c^2_g = 1$, which is the only observationally important speed in this limit. It is very important to note that in such a singly coupled limit $r\rightarrow 0$ or $r\rightarrow \infty$ are not necessarily dangerous since the potentially singular terms containing $\frac{1}{r}$ (as well as the terms containing $r$, which are dangerous when $r\rightarrow\infty$) are multiplied by both $\alpha$ and $\beta$ and therefore vanish in either the case of $\alpha = 0$ or $\beta = 0$. Putting all these discussions together we arrive at an important statement: the propagation of gravitational waves in doubly coupled bigravity is viable {\it if and only if} $b = 1$ or we are in a singly coupled regime.

It is important to note that the current bounds on the speed of GWs have been placed through the observations at very low redshifts ($z\approx 0$), i.e. at almost the present time. This means that, strictly speaking, the viability conditions we discussed above are required to hold only at $z\approx 0$, including the condition $b=1$. Let us for now assume that the constraint on the speed of GWs is valid not only in the present epoch but it applies also to the earlier epochs of the Universe, i.e. we assume $b = 1$ at all times. Later on, when we discuss our numerical analysis, we show a rather vigorous feature of the theory that imposing $b|_{z\approx0}=1$ will force $b$ to be unity at all redshifts.

Imposing $b(z) = 1$ at all times tells us that the two background metrics $g_{\mu\nu}$ and $f_{\mu\nu}$ should be proportional. This can easily be seen by setting $b(z)=1$ in Eq. (\ref{eq:bdef}) and noting that $r=a_f/a_g$, resulting in
\begin{equation}
\frac{a_f(z)}{a_g(z)} = C = \frac{N_f(z)}{N_g(z)}\,,
\end{equation}
with $C$ being some (constant) proportionality factor. In order to understand under which circumstances these proportional solutions exist, let us consider the early-time and late-time asymptotic limits of Eq. (\ref{eq:algebraic_constraint}). By taking the future asymptotic limit, with $\rho \rightarrow 0$, we obtain
\begin{equation}
\beta_3 r_\infty^4\!+\!(3\beta_2\!-\!\beta_4)r_\infty^3\!+\!3 (\beta_1\!-\!\beta_3) r_\infty^2\!+\!(\beta_0\!-\!3\beta_2)r_\infty\!-\!\beta_1\!=\!0 \,
\label{prop_metric_equation_y}
\end{equation}
for the value of $r$ in the far future, $r_\infty$. Note that $r_\infty$ being a solution of this time-independent equation means that it is a constant. This in turn means that the two metrics are necessarily proportional in the far-future limit. Additionally, the early-Universe limit of Eq. (\ref{eq:algebraic_constraint}) fixes the value of $r$ to either $\gamma$ or $-\gamma$. The latter does not give viable cosmologies~\cite{Enander:2014xga}, and therefore $r\rightarrow r_{-\infty}=\gamma$ is the only viable early-time limit. Restricting to the solutions for which $r$ does not exhibit any singular behavior~\cite{Enander:2014xga}, one can show that $r$ should {\it monotonically} evolve between $r=r_{-\infty}$ and $r=r_\infty$ over the history. The monotonicity of $r$ implies that when the two limiting values $r_{-\infty}$ and $r_{\infty}$ coincide, i.e. when $r_\infty=\gamma$, we have constant $r$ over the entire history of the Universe and hence the background metrics are proportional in that case.
\begin{figure}[h!]
\center
  \includegraphics[height=5cm]{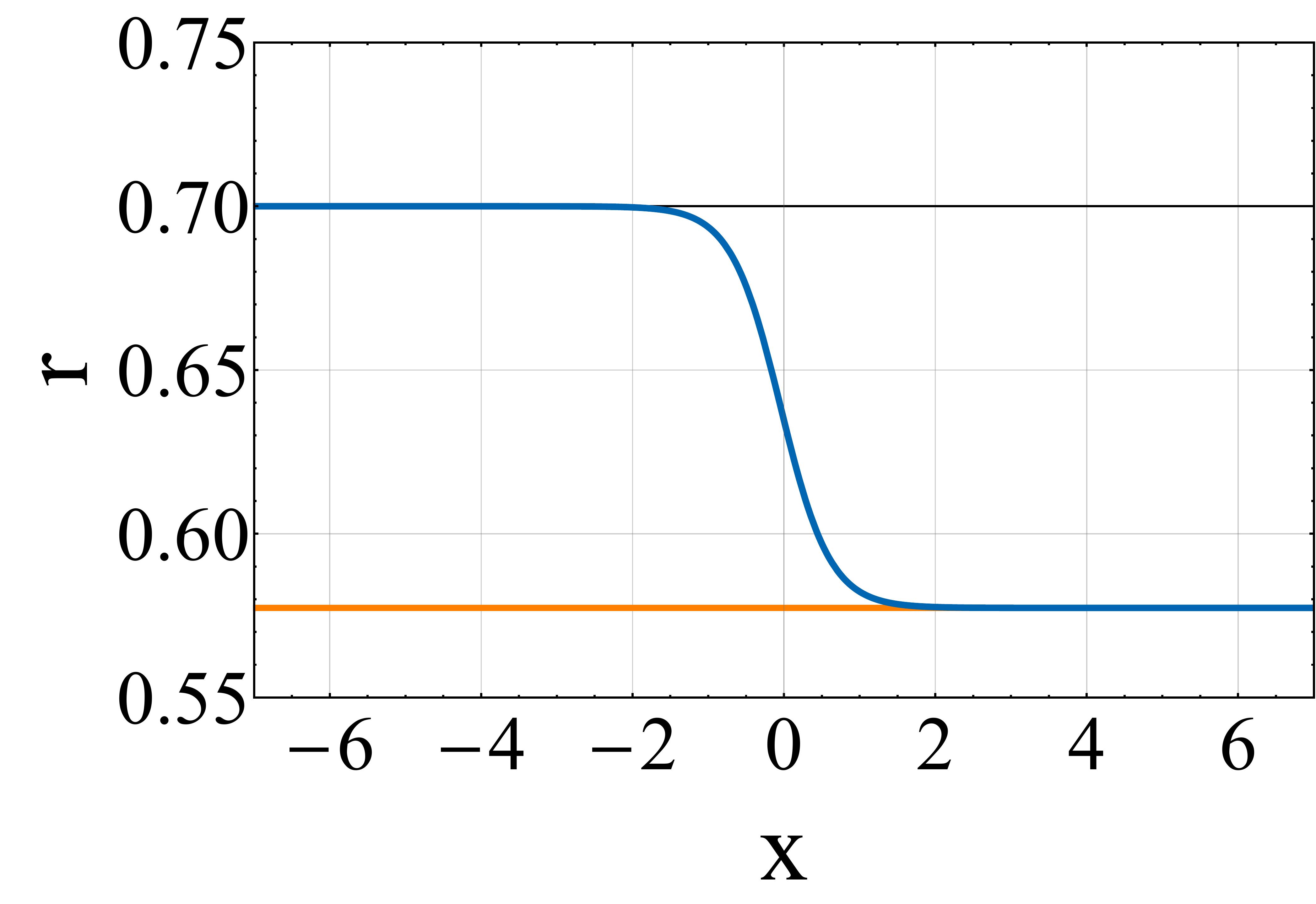}
\caption{\footnotesize \label{fig:r_plot}Behavior of $r$, the ratio of the scale factors of the two metrics, as a function of the number of $e$-folds $x$, with $x=0$ corresponding to the present time. The evolution of $r$ has been shown with (thick) blue and orange curves for two different values of $\gamma$, both for a single-interaction-parameter model with only $\beta_1$ being turned on. The blue curve corresponds to a case where $\gamma$ does not satisfy the special tuning condition for proportional metrics. The curve exhibits two constant-$r$ epochs of $r_{-\infty} = \gamma$ and $r_{\infty} = 1/\sqrt{3}$, with the latter being the solution of Eq. (\ref{prop_metric_equation_y}) regardless of the value of $\beta_1$. The orange curve corresponds to a case where $\gamma$ is chosen such that it is the solution of Eq. (\ref{prop_metric_equation_y}), i.e. $\gamma = r_{\infty} = 1/\sqrt{3}$.}
\end{figure}

Based on the discussions above, we can now formulate the necessary and sufficient conditions for the two background metrics to be proportional:
\begin{enumerate}
\item Background solutions are proportional {\it if and only if} $r$  is given by $r = \gamma$ at all times, where $\gamma \equiv \beta/\alpha$. Note that one does not need to check whether this condition holds at all times; as we argued above, because of the monotonicity of $r$, having $r=\gamma$ even at one instant in time, other than the asymptotic past, is sufficient for the condition to be satisfied at all times.
\item Equivalently, the background solutions are proportional {\it if and only if} the parameters of the model solve the algebraic equation
\begin{equation}
\beta_3\gamma^4\!+\!(3\beta_2\!-\!\beta_4)\gamma^3\!+\!3 (\beta_1\!-\!\beta_3) \gamma^2\!+\!(\beta_0\!-\!3\beta_2)\gamma\!-\!\beta_1\!=\!0 \,.
\label{prop_metric_equation}
\end{equation}
\end{enumerate}

We demonstrate these conditions in Fig.~\ref{fig:r_plot} by plotting the dependence of $r$ on the number of $e$-folds $x$, with the present time given by $x = 0$, for a single-interaction-parameter scenario where only $\beta_1$ is turned on while $\beta_{0,2,3,4}=0$. The blue curve corresponds to a case where $\gamma$ does not satisfy the special tuning condition for proportional metrics. The curve exhibits two constant-$r$ epochs. The far-past epoch corresponds to $r = \gamma$ (the horizontal, thin, black line), while the far-future limit is given by the solution of Eq. (\ref{prop_metric_equation_y}) for which $r_{\infty} = 1/\sqrt{3}$ regardless of the value of $\beta_1$. The orange curve corresponds to a case where $\gamma$ is chosen such that it is the solution of Eq. (\ref{prop_metric_equation_y}), i.e. $\gamma = r_{\infty} = 1/\sqrt{3}$. The value of $\beta_1$ is not relevant for the arguments here because in this case the asymptotic value $r_{\infty}$ is independent of the value of $\beta_1$ (the value of $r_{-\infty}$ is always independent of the values of $\beta_n$ parameters). In order to illustrate  our arguments, we have chosen two different values of $\beta_1$ for producing the two curves (blue and orange). As expected, they agree in the far-future limit, even though the values of $\beta_1$ are different for the two curves.

As we see in the next section, bigravity models for which only one of the $\beta_{0,1,2,3,4}$ parameters is turned on are particularly interesting. For those cases the proportional background solutions correspond to the following values of the parameter $\gamma$:
\begin{enumerate}
\item $\beta_0$ or $\beta_4$ only: $\gamma = r_{\infty} = 0$\,,
\item $\beta_1$ only: $\gamma = r_{\infty} = \frac{1}{\sqrt{3}}$\,,
\item $\beta_2$ only: $\gamma = r_{\infty} = 1$\,,
\item $\beta_3$ only: $\gamma = r_{\infty} = \sqrt{3}$\,.
\end{enumerate}
Note that $\gamma$ and therefore $r_{\infty}$ in these cases are independent of the value of the corresponding $\beta_n$ parameter. Note also that, as we discussed in the previous section, the single-parameter models with only $\beta_1$ or $\beta_3$ turned on are identical, as long as $r\leftrightarrow1/r$ (or equivalently $\gamma\leftrightarrow 1/\gamma$), justifying the values $1/\sqrt{3}$ and $\sqrt{3}$ for $r_\infty$ in these models. In addition, it is interesting to notice that for the $\beta_0$ and $\beta_4$ only models, proportional backgrounds do not exist, as in those cases $\gamma$ is forced to be vanishing, and therefore the theory becomes singly coupled.

All these cases of proportional background metrics with only one of the $\beta_{1,2,3}$ parameters being nonzero can be verified easily by applying the Bianchi constraint $H_g=rH_f$ to the Friedmann equations (\ref{eq:friedg}) and (\ref{eq:friedf}), obtaining
\begin{align}
3H^2_g &= \frac{1}{M^2_{\text{eff}}}\rho(1 + \gamma r)^3+H^2_0(\beta_0+3\beta_1r+3\beta_2r^2 + \beta_3r^3)\,,\label{eq:friedmanng}\\
3H^2_g &= \frac{\gamma}{M^2_{\text{eff}}}\rho\frac{(1 + \gamma r)^3}{r}+H^2_0(\frac{\beta_1}{r}+3\beta_2+3\beta_3r+\beta_4r^2)\,.\label{eq:friedmannf}
\end{align}
In general, we have two dynamical variables $a_g$ and $a_f$, which are determined by the two independent, dynamical equations (\ref{eq:friedmanng}) and (\ref{eq:friedmannf}). Now, if the two metrics are proportional, this  means that $a_g$ and $a_f$ are also proportional, and $r$ is a constant. We then have effectively only one dynamical variable, $a_g$ or $a_f$, and the two dynamical equations (\ref{eq:friedmanng}) and (\ref{eq:friedmannf}) must be identical. This means that the right-hand sides of the two equations should be identically the same. Now, setting all the parameters $\beta_n$ to $0$, except for either of $\beta_1$, $\beta_2$, or $\beta_3$, we immediately arrive at the values for $r_\infty$ and $\gamma$ presented above for these three cases.

Now turning back to the condition for the speed of the gravitational waves to be identical to the speed of light, we argued that what is strictly needed is to have $b|_{z\approx0} \approx 1$, as the speed of GWs has been measured only at the present epoch $z\approx0$. If, additionally, the parameters of the model giving $b|_{z=0} = 1$ satisfy the algebraic equation (\ref{prop_metric_equation}) then they lead to proportional background solutions and the $b = 1$ condition is satisfied at all times, implying necessarily that $c_g = c_f = 1$ at all times. The question of whether a set of parameters giving $b|_{z=0} = 1$ (hence $c_g|_{z=0} = c_f|_{z=0} = 1$) while not satisfying Eq. (\ref{prop_metric_equation}) can happen in our doubly coupled bigravity models cannot be answered based on our analytical arguments here, and needs a numerical scanning of the parameter space. In principle it could be possible that the two background metrics are not proportional while $b$ becomes unity at the present epoch simply as a coincidence for a specific combination of the parameters. We however demonstrate later that for all the models that we study in this paper the cosmologically viable solutions with $b|_{z=0} = 1$ also satisfy Eq. (\ref{prop_metric_equation}), implying $b = 1$ at all times, and therefore the proportionality of the background metrics.

\section{MCMC scans and observational constraints}\label{constraints}

In this section we present the results of a set of MCMC scans of the parameter space of doubly coupled bigravity when different sets of  parameters are allowed to vary while the rest are fixed to $0$. We should first emphasize that we do not intend here to perform a detailed parameter estimation of the model using  cosmological observations. This has been done in Ref.~\cite{Enander:2014xga} using the geometrical constraints on cosmic histories at the background level.\footnote{Note, however, that the MCMC scans presented in Ref.~\cite{Enander:2014xga} include only single-$\beta_n$ models, while in the current paper we consider also the cosmological constraints on two-parameter models.} We are rather interested in studying the impact of the constraints from the measurements of gravitational waves and the bounds on their speed on the cosmologically viable regions of the parameter space. We first perform MCMC scans of the models using similar cosmological data sets as those used in Ref.~\cite{Enander:2014xga}. The geometrical constraints that we consider are a combination of the observed angular scales of the cosmic microwave background anisotropies~\cite{Ade:2015xua}, the supernovae redshift-luminosity relation~\cite{Betoule:2014frx}, the measurements of the baryon acoustic oscillations (BAO)~\cite{Beutler:2011hx,Blake:2011en,Anderson:2012sa,Anderson:2013zyy,Ross:2014qpa}, and the local measurement of the Hubble constant $H_{0}$~\cite{Riess:2016jrr}. Our scans provide a set of points in the parameter space of the models all of which are in good agreement with cosmological observations. We have checked that our results are in perfect agreement with the results of Ref.~\cite{Enander:2014xga} for the cases studied in that paper. We then explore the implications of imposing the GW constraints on the points, and investigate whether and how strongly the cosmologically viable regions are affected by the GW observations.

Our full bigravity model contains seven free parameters, as far as our MCMC scans are concerned. These include the five $\beta_n$ parameters for the interaction terms, the ratio of the couplings of the two metrics to matter $\gamma$, and the present value of the matter density parameter $\Omega_\text{M}^{0}$, defined as
\begin{equation}
\Omega_\text{M}^{0}\equiv\frac{\rho_\text{M}^{0}}{3M_\text{eff}^2H_0^2} \,.
\end{equation}
Note that one should not necessarily expect to obtain a value for $\Omega_\text{M}^{0}$ similar to the best-fit one in the standard model of cosmology, $\Lambda$CDM, for a bigravity model that fits the data well, even for proportional backgrounds where the interaction terms contribute with a $\Lambda$-like constant to the Friedmann equation. The reason, as explained in Ref.~\cite{Enander:2014xga} in detail, is the extra factor appearing in the matter density term of the Friedmann equation. We see below that indeed in some cases the viable points in the parameter space give values for $\Omega_\text{M}^{0}$ that are significantly smaller than the $\Lambda$CDM value of $\sim0.3$.

For each point in the parameter space of the theory we also output the corresponding values of $r$, $b$, $c_g$ and $c_f$, all evaluated at the present time. These allow us to check which parts of the parameter space agree with the observational constraint $c_{g}\approx 1$ (or $c_{f}\approx 1$), and to verify explicitly the conditions on $b$ and $r$. We particularly use the quantity $(c_g^2-1)(c_f^2-1)$ as a measure of how fit a point is to the observational constraints on the speed of GWs.

We perform our MCMC scans for various submodels, namely the single-parameter\footnote{This is only a terminological convention here, and strictly speaking, our single-parameter models have two free parameters, as $\gamma$ is always a free parameter of the models.} models of $\beta_0$, $\beta_1$, and $\beta_2$ (with other $\beta_n$ being set to zero in each case), and the two-parameter models of $\beta_0\beta_1$, $\beta_0\beta_2$, $\beta_1\beta_2$, and $\beta_1\beta_3$. One should note that, as we discussed before, the single-parameter models of $\beta_3$ and $\beta_4$ are identical to the $\beta_1$ and $\beta_0$ models, respectively, because of the duality properties of the theory. In addition, for the same reason, each one of the other two-parameter models is equivalent to one of the two-parameter models considered here, and their phenomenologies are therefore already captured. Our objective in this paper is not to perform a detailed and extensive statistical analysis of the entire parameter space of  doubly coupled bigravity, and we are mainly interested in a qualitative understanding of the implications of the GW observations for the viability of the theory, which can very well be captured in the studies of single-parameter and two-parameter cases. We therefore do not discuss three- or higher-parameter models. As we see later, although the constraints are quite strong for most of these cases, the parameter space in some models still allows viable cosmologies, and clearly, by increasing the number of free parameters one expects to enlarge the number of possibilities for finding viable scenarios within the model. We leave a detailed statistical analysis of the full model for future work.

\subsection{One-parameter models}

\begin{figure}[h!]
\center
  \includegraphics[height=3cm]{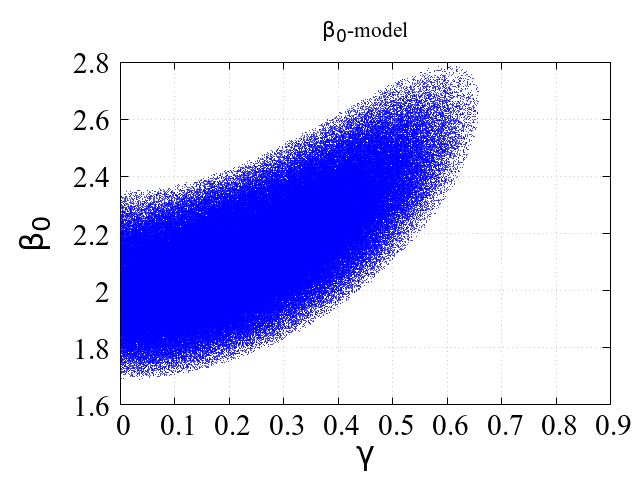}
  \includegraphics[height=3cm]{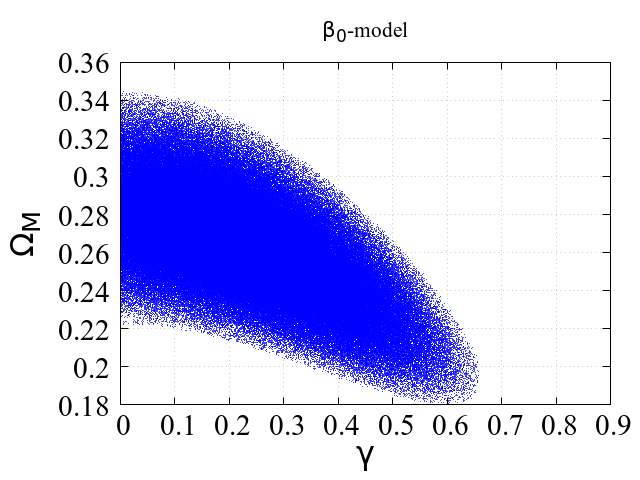}
  \includegraphics[height=3cm]{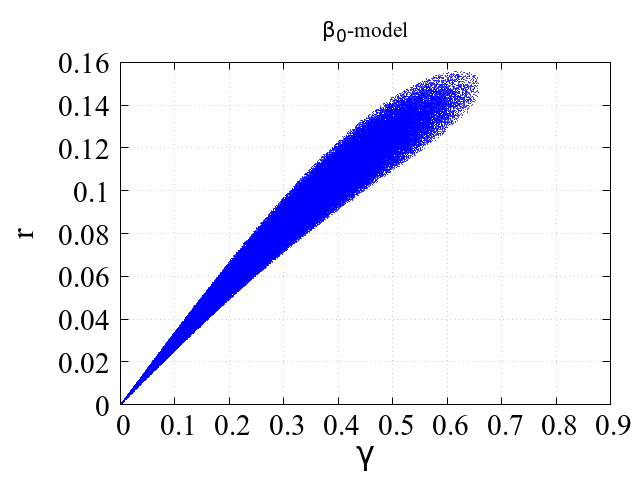}
  \includegraphics[height=3cm]{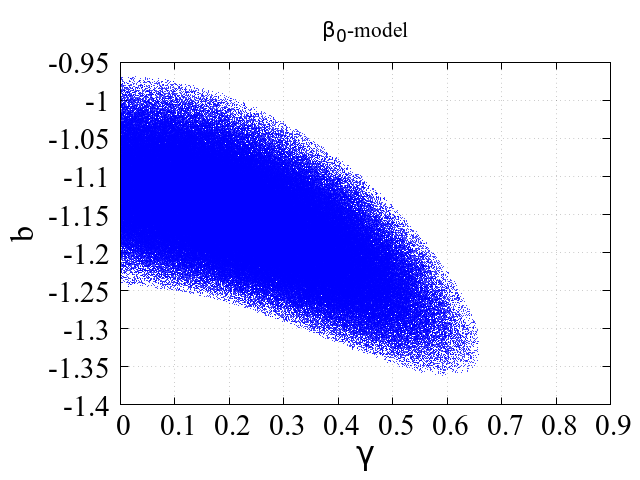}
  \includegraphics[height=3cm]{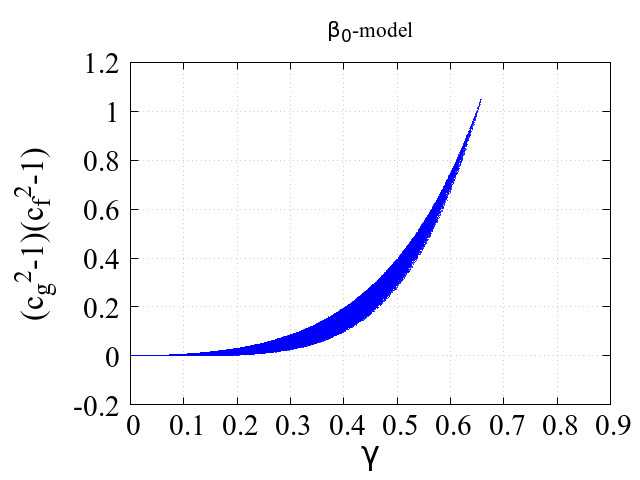}
\caption{\footnotesize\label{fig:beta0_plot1} Scatter plots showing all the cosmologically viable points in the parameter space of the doubly coupled $\beta_0$ model, where all the interaction parameters $\beta_n$ are set to $0$ except for $\beta_0$, which is allowed to vary. The plots show the constraints on $\beta_0$, $\Omega_\text{M}^{0}$, $r$ (the ratio of the scale factors of the two metrics $g_{\mu\nu}$ and $f_{\mu\nu}$), $b\equiv\frac{1}{r}\frac{N_f}{N_g}$, and the quantity $(c_g^2-1)(c_f^2-1)$ (capturing the deviations of the $g$ and $f$ gravitational wave speeds from the speed of light), all versus $\gamma\equiv\frac{\beta}{\alpha}$. Note that $c_{g}$, $c_{f}$, $b$, and $r$ are all computed at $z=0$, i.e. at the present time. In this $\beta_0$ model, the only part of the parameter space that is left after imposing $c_{g}=1$ or $c_{g}=1$ is the singly coupled submodel characterized by $\gamma=0$.}
\end{figure}

{\bf $\bullet$ $\pmb{\beta_0}$ model}: Let us first emphasize that, contrary to singly coupled bigravity, in the doubly coupled theory the parameters $\beta_0$ and $\beta_4$ are no longer the explicit cosmological constants corresponding to the two metrics $g_{\mu\nu}$ and $f_{\mu\nu}$. The reason is that matter couples to the effective metric $g_{\mu\nu}^\text{eff}$, which is a combination of $g_{\mu\nu}$ and $f_{\mu\nu}$. This can be seen explicitly by looking at the effective Friedmann equation (\ref{eq:friedeff}) and comparing it with Eqs.~(\ref{eq:friedg}) and (\ref{eq:friedf}). In addition, in the singly coupled theory, where matter couples to, say, $g_{\mu\nu}$, $\beta_0$ behaves as the matter vacuum energy in the action of the theory, as it appears in the interaction terms as $\beta_0\sqrt{-g}$ (note that $e_0=1$). In the doubly coupled theory, however, all the interaction parameters $\beta_n$ directly receive contributions from quantum matter loops, and the definition of vacuum energy is more subtle than in the singly coupled theory. It is therefore interesting to study a single-parameter, doubly coupled model with only $\beta_0$ turned on, while all the other parameters $\beta_n$ are set to $0$: for the singly coupled case this is nothing but $\Lambda$CDM. The cosmology of this $\beta_0$ model in doubly coupled bigravity has been studied in Ref.~\cite{Enander:2014xga}. We reproduce and show the cosmological constraints on the three parameters $\beta_{0}$, $\Omega_\text{M}^{0}$, and $\gamma$ in the upper panels of Fig.~\ref{fig:beta0_plot1}, which are in full agreement with the results of Ref.~\cite{Enander:2014xga}. Note that $\gamma=0$ corresponds to the singly coupled scenario, which reduces to $\Lambda$CDM for this $\beta_0$-only model.

Let us now look at the lowest panel of Fig.~\ref{fig:beta0_plot1}, where the present value of $(c_g^2-1)(c_f^2-1)$ has been depicted versus $\gamma$. This plot shows that in order for the model to be cosmologically viable and simultaneously predict gravitational waves with the speed equal to the speed of light (i.e. for at least one of the two quantities $c_{g}$ and $c_{f}$ to be unity), $\gamma$ is required to be $0$, which in turn implies that the model needs to be singly coupled. In this case $r$ is forced to be vanishing, although $r$ is no longer a meaningful quantity as there is no interaction between $g_{\mu\nu}$ and $f_{\mu\nu}$, and $f_{\mu\nu}$ completely decouples from the theory. This all tells us that the $\beta_{0}$ model satisfies the cosmological and gravitational wave constraints only in its singly coupled limit, which is equivalent to $\Lambda$CDM. We do not see any cases of proportional metrics in this model, as such cases should also give GWs consistent with observations. Let us take a closer look at this and understand why such a situation does not happen in the $\beta_{0}$ model by looking again at the condition for proportional background metrics. As we argued in the previous section, for proportional backgrounds $\gamma$ must satisfy Eq. (\ref{prop_metric_equation}), while $r_{\infty}=\gamma$. Setting all $\beta_n$ parameters to $0$ except for $\beta_0$, we arrive at $\gamma=r_{\infty} = 0$. First of all, this is exactly what we see in the middle, left panel of Fig.~\ref{fig:beta0_plot1} for $r$ and $\gamma$. Additionally, we are back to the condition $\gamma=0$ that corresponds to a single coupling. This means that the $\beta_{0}$ model does not admit any sets of (nontrivial) proportional backgrounds, unless we consider $f_{\mu\nu}$ to be proportional to $g_{\mu\nu}$ with a vanishing proportionality factor. The fact that this is a peculiar case can also be seen by looking at the middle, right panel of Fig.~\ref{fig:beta0_plot1}, which shows $b$ versus $\gamma$. $b$ is always negative, which means that the condition for proportional backgrounds, $b=1$, can never be satisfied.

\begin{figure}[h!]
\center
  \includegraphics[height=3cm]{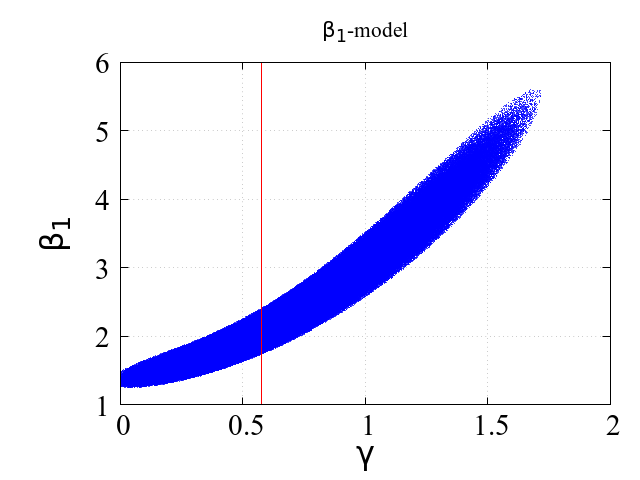}
  \includegraphics[height=3cm]{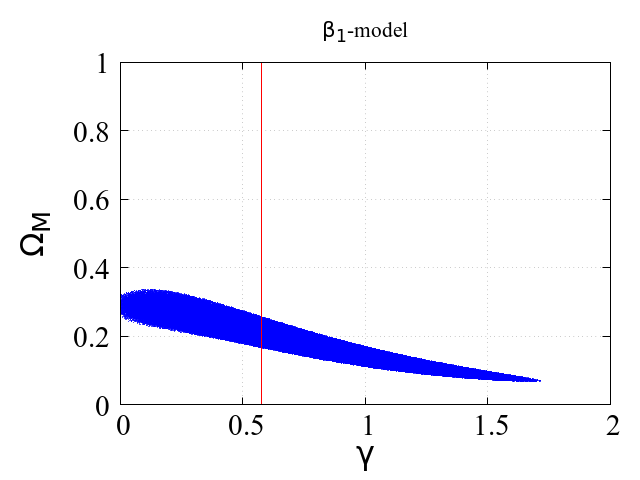}
  \includegraphics[height=3cm]{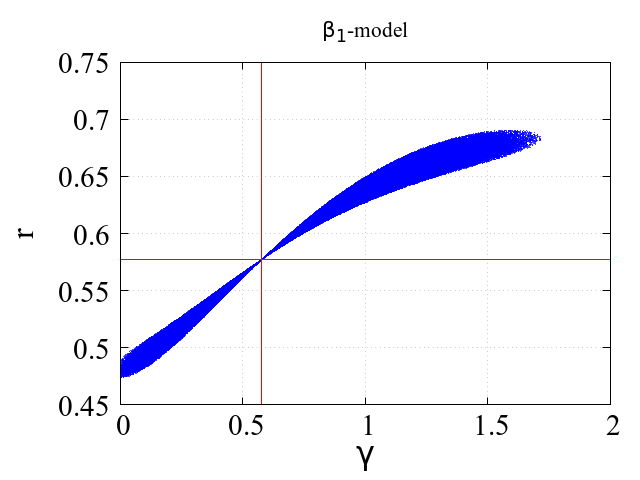}
  \includegraphics[height=3cm]{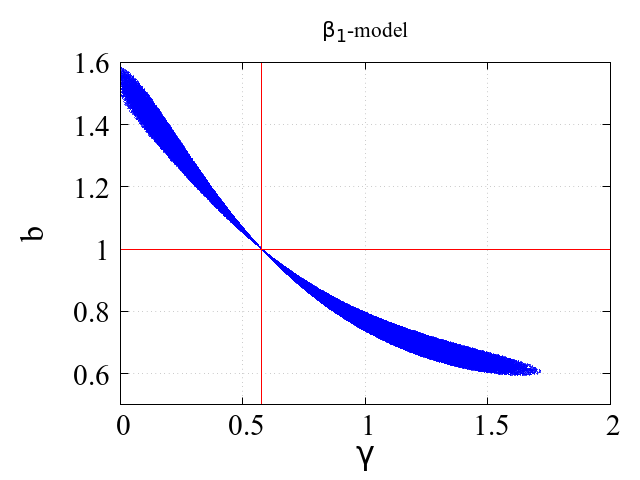}
  \includegraphics[height=3cm]{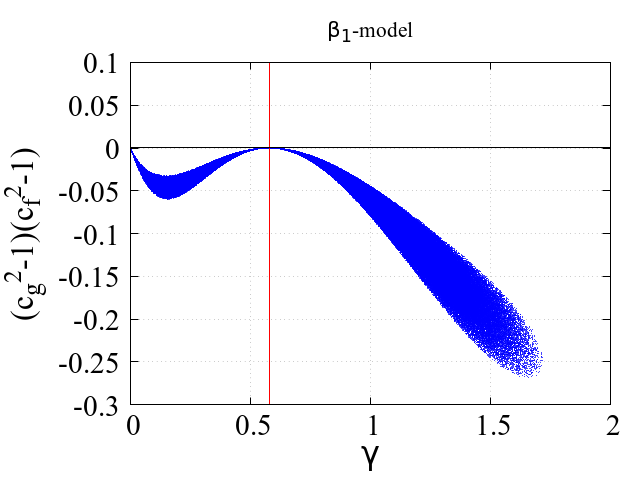}
\caption{\footnotesize\label{fig:beta1_plot1} The same as in Fig.~\ref{fig:beta0_plot1}, but for the doubly coupled $\beta_1$ model where all interaction parameters $\beta_n$ are set to $0$ except for $\beta_1$. In this case, the only parts of the parameter space that are left after imposing $(c_g^2-1)(c_f^2-1)=0$ are the singly coupled submodel characterized by $\gamma=0$, and the solutions with the two background metrics being proportional, with $\gamma=1/\sqrt{3}$, illustrated by the red lines in the plots.}
\end{figure}

{\bf $\bullet$ $\pmb{\beta_1}$ model}: Here we turn on only the $\beta_1$ parameter and set to $0$ all the other interaction parameters $\beta_{0,2,3,4}$. From our discussions in the previous section, we expect this submodel to give the speed of gravity waves equal to the speed of light for the cases with $r_\infty = \gamma = 1/\sqrt{3}$, where the background metrics are proportional, as well as for the singly coupled corners with $\gamma=0$. The lowest panel of Fig.~\ref{fig:beta1_plot1} presents the dependence of $(c_g^2-1)(c_f^2-1)|_{z=0}$ on the value of $\gamma$ as a result of our numerical scans. We first notice that no viable combinations of the parameters provide $c_g$ and $c_f$ both larger or smaller than the speed of light, as $(c_g^2-1)(c_f^2-1)$ is always negative or $0$. The plot also shows two points with $(c_g^2-1)(c_f^2-1)=0$, one of which being the obvious limit of single coupling with $\gamma = 0$, and the other one, as expected, corresponding to the case of proportional backgrounds with $\gamma = 1/\sqrt{3}$, depicted by the vertical, red line. This becomes more clear by looking at the middle panels of Fig.~\ref{fig:beta1_plot1}, showing $r$ and $b$ versus $\gamma$. The red lines in the plots show that indeed $\gamma = 1/\sqrt{3}$ corresponds to $r = 1/\sqrt{3}$ and $b=1$, as expected. Also note that $b$ is always positive for all the cosmologically viable points in the parameter space of this model. Finally, the upper panels of Fig.~\ref{fig:beta1_plot1} show the constraints on $\beta_1$ and $\Omega_\text{M}^{0}$ versus $\gamma$, with the vertical lines again showing the condition for the two background metrics to be proportional, with $\gamma = 1/\sqrt{3}$ giving $(c_g^2-1)(c_f^2-1)=0$: all the points residing on the lines are viable. Although most of the original, cosmologically viable points are now excluded and the model is highly constrained, our results show that there still remains some freedom in choosing $\beta_1$ for the fixed $\gamma = 1/\sqrt{3}$. It is also interesting to note that the preferred values of $\Omega_\text{M}^{0}$ are smaller than the $\Lambda$CDM value of $\sim0.3$. In summary, as expected, the viable points in the parameter space of the model correspond to the scenarios which do not represent the full dynamics of the doubly coupled model. One remaining region is the singly coupled limit, and the other one corresponds to the cases where the background metrics are proportional, and we again effectively have only one dynamical metric at work. In this latter case, the model is effectively equivalent to $\Lambda$CDM, at the level of the background (and linear perturbations~\cite{Schmidt-May:2014xla}).

\begin{figure}[h!]
\center
  \includegraphics[height=3cm]{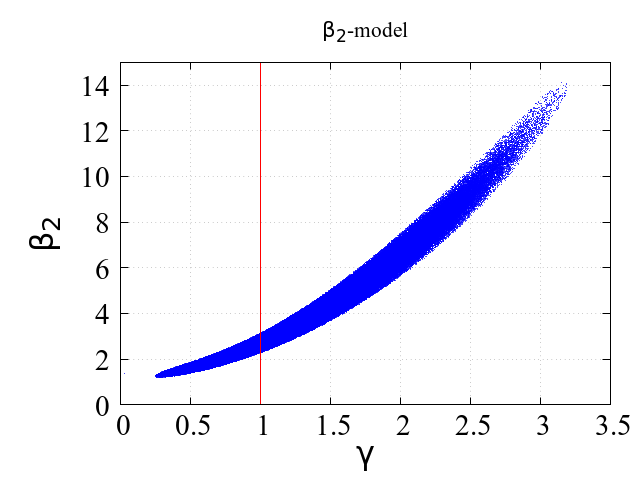}
  \includegraphics[height=3cm]{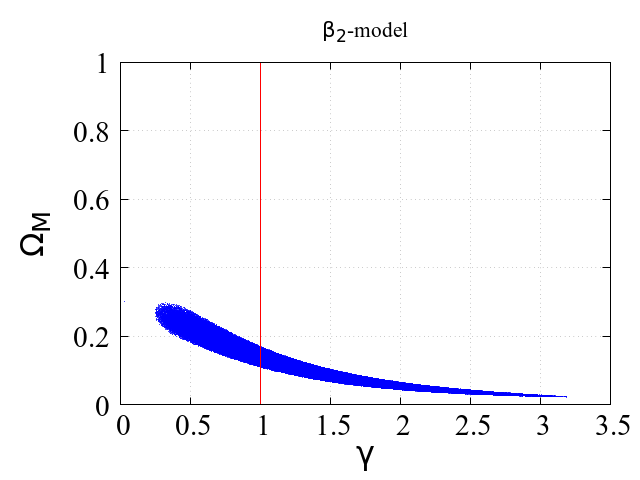}
  \includegraphics[height=3cm]{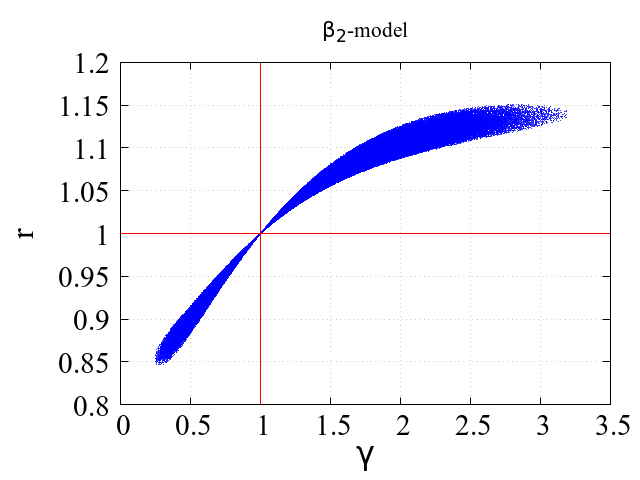}
  \includegraphics[height=3cm]{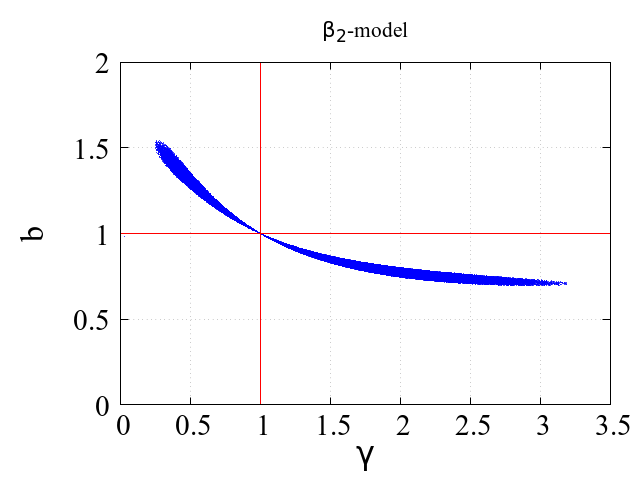}
  \includegraphics[height=3cm]{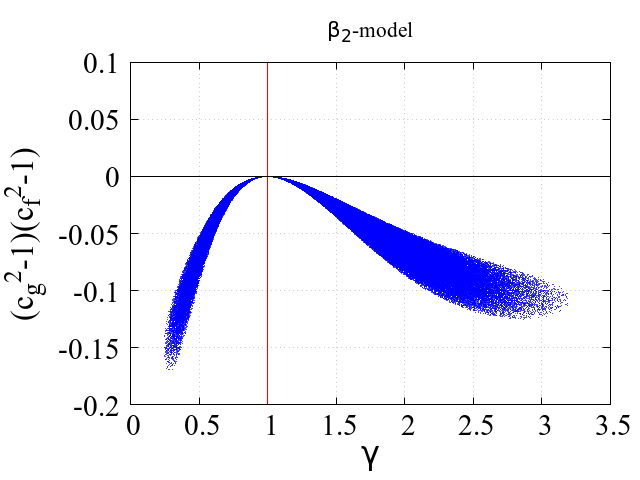}
\caption{\footnotesize\label{fig:beta2_plot1} The same as in Figs.~\ref{fig:beta0_plot1} and~\ref{fig:beta1_plot1}, but for the doubly coupled $\beta_2$ model where all interaction parameters $\beta_n$ are set to $0$ except for $\beta_2$. In this case, the only part of the parameter space consistent with $(c_g^2-1)(c_f^2-1)=0$ is the one corresponding to the two background metrics being proportional, with $\gamma=1$.}
\end{figure}

{\bf $\bullet$ $\pmb{\beta_2}$ model}: Fig.~\ref{fig:beta2_plot1} presents the results of our MCMC scans for the model with only $\beta_2$ turned on. All the panels clearly show that the singly coupled subset of the parameter space (with $\gamma=0$) is not viable cosmologically as there are no points with $\gamma=0$ that fit the data. This is in agreement with the results of Ref.~\cite{Akrami:2013pna}. The model, however, provides excellent fits to the data for $\gamma\gtrsim0.3$. Looking now at the lowest panel of Fig.~\ref{fig:beta2_plot1}, we see that the only points in the parameter space that are consistent with $(c_g^2-1)(c_f^2-1)=0$ today, i.e. with the bounds from the GW observations, are the ones for which $\gamma=1$, meaning that the metrics are proportional. These points correspond to $b=1$ (see the middle, right panel). This is in agreement with our findings in the previous section for the $\beta_2$ model, with $r_{\infty}=\gamma=1$ for proportional metrics. For all the other cosmologically viable points the tensor modes of one of the two metrics $g_{\mu\nu}$ and $f_{\mu\nu}$ travel faster and the other ones travel slower than light. Finally, the upper panels show the constraints on the model parameters $\beta_2$ and $\Omega_\text{M}^{0}$, with again lower preferred values for $\Omega_\text{M}^{0}$ compared to $\Lambda$CDM.

\subsection{Two-parameter models}

Let us now turn on two of the interaction parameters $\beta_n$ and let them vary. As we argued earlier, many of these submodels are physically equivalent because of the symmetry of the theory. We therefore study four representative cases of $\beta_0\beta_1$, $\beta_0\beta_2$, $\beta_1\beta_2$, and $\beta_1\beta_3$ models. Note that even though for example the model with only $\beta_1$ turned on is identical to the model with only $\beta_3$ turned on, when the two parameters are both nonzero the resulting two-parameter model can in general be very different from the single-parameter ones, with generally richer phenomenologies. The reason is that the two parameters can take two different values, making the model different from the cases with only one of the parameters left free.

\begin{figure*}
\center
  \includegraphics[height=5.5cm]{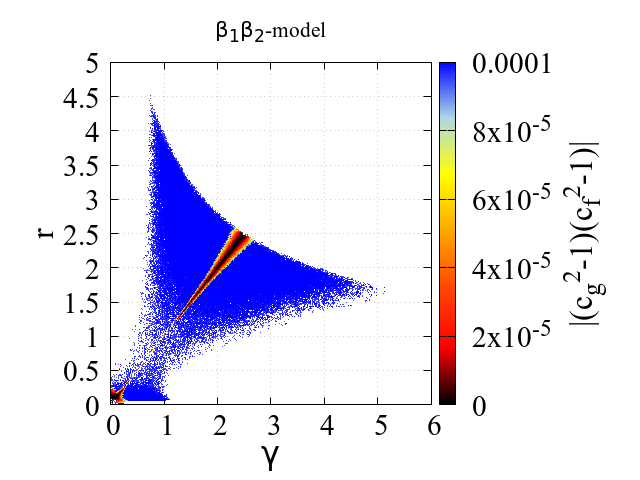}
  \includegraphics[height=5.5cm]{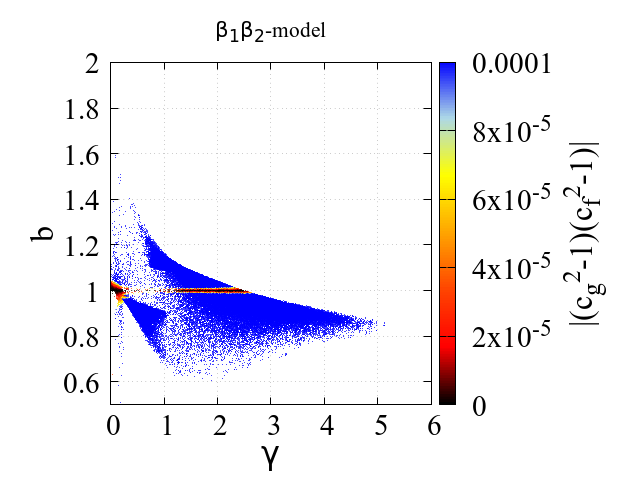}
  \includegraphics[height=5.5cm]{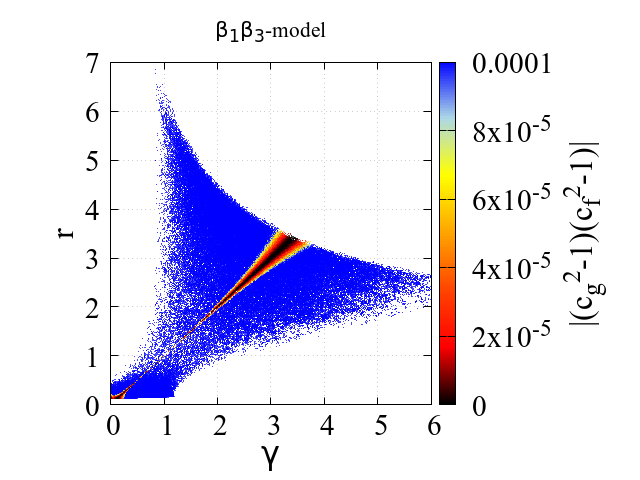}
  \includegraphics[height=5.5cm]{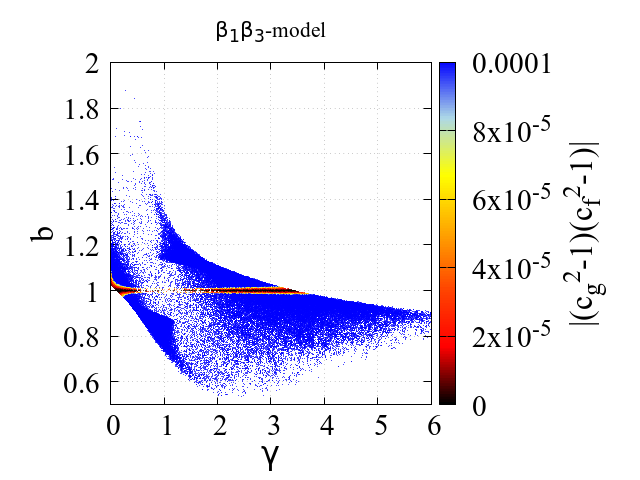}
  \includegraphics[height=5.5cm]{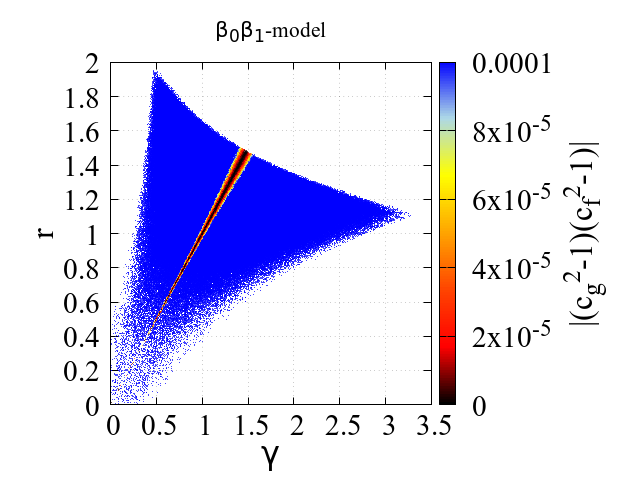}
  \includegraphics[height=5.5cm]{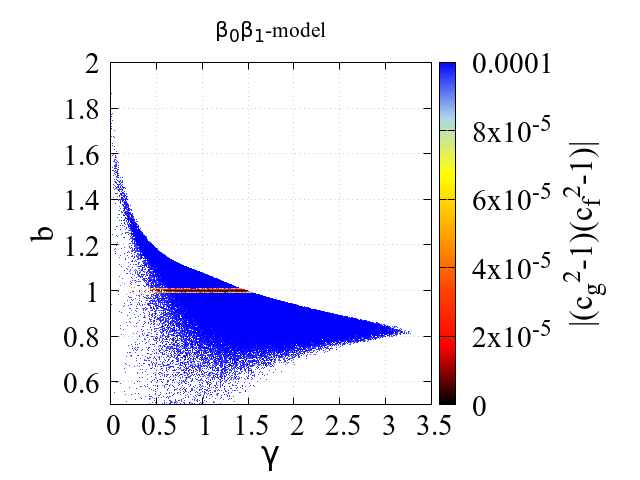}
  \includegraphics[height=5.5cm]{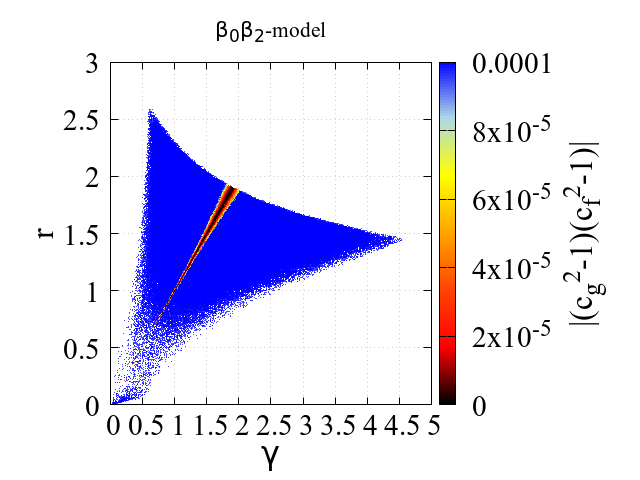}
  \includegraphics[height=5.5cm]{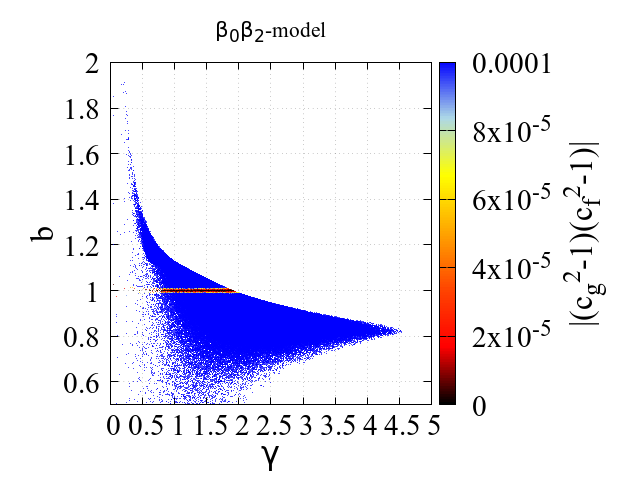}
\caption{\footnotesize\label{fig:twoparam_plot1} Results of the MCMC scans for the two-parameter models $\beta_1\beta_2$, $\beta_1\beta_3$, $\beta_0\beta_1$, and $\beta_0\beta_2$. All the cosmologically viable points are shown in the $r-\gamma$ and $b-\gamma$ planes, and the color in each panel shows the values of $|(c_g^2-1)(c_f^2-1)|$ as a measure for how fit the points are to the bounds on the speed of gravitational waves. Here, $r$, $b$, and $|(c_g^2-1)(c_f^2-1)|$ are all computed at the present time ($z=0$).}
\end{figure*}

The results of our MCMC scans for these models are presented in Fig.~\ref{fig:twoparam_plot1}, where the quantities $r$ and $b$ (both computed at the present time) are given in terms of the coupling ratio $\gamma$. The color code shows the values of $|(c_g^2-1)(c_f^2-1)|$.

\begin{figure}
\center
  \includegraphics[height=3.2cm]{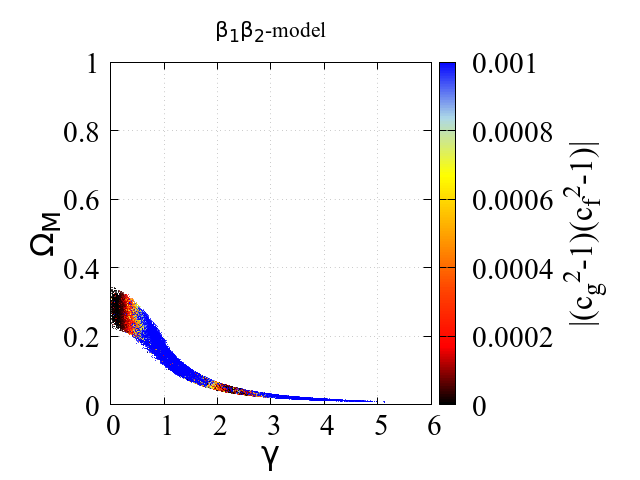}
  \includegraphics[height=3.2cm]{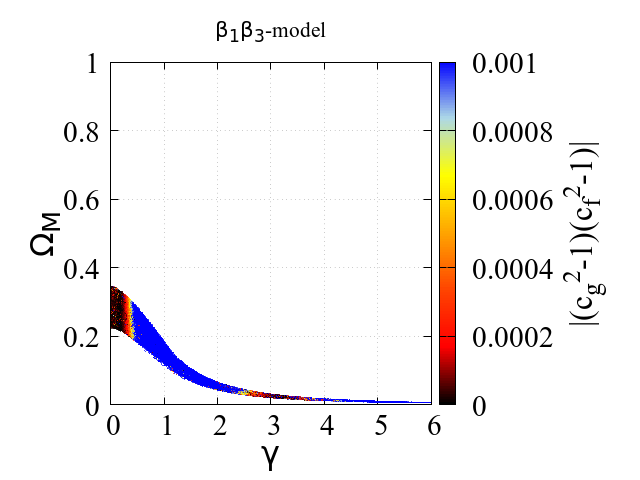}
  \includegraphics[height=3.2cm]{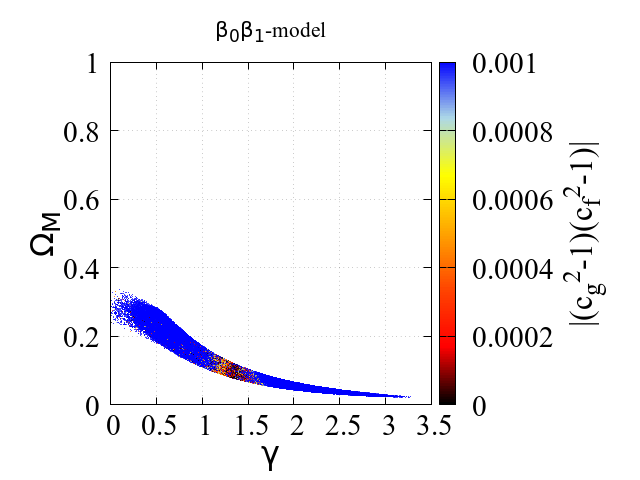}
  \includegraphics[height=3.2cm]{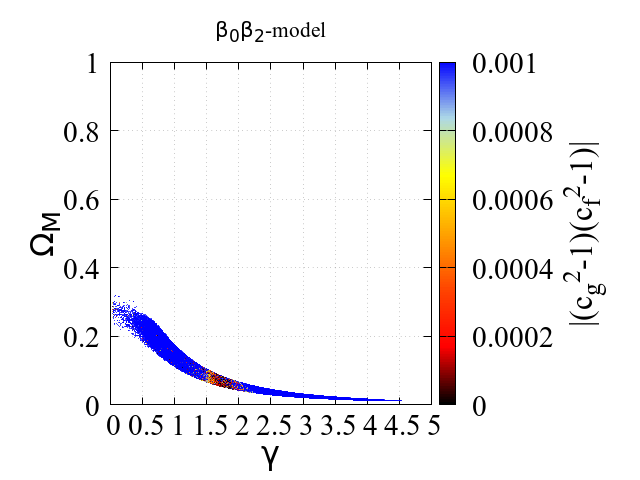}
\caption{\footnotesize\label{fig:twoparam_plot2} Constraints on $\Omega_\text{M}^{0}$, the present value of the matter density parameter, for the two-parameter models $\beta_1\beta_2$, $\beta_1\beta_3$, $\beta_0\beta_1$, and $\beta_0\beta_2$. All the cosmologically viable points are shown and the color in each panel shows the values of $|(c_g^2-1)(c_f^2-1)|$ as a measure for how fit the points are to the bounds on the speed of gravitational waves. Note that for the proportional backgrounds (i.e. the favored regions in the plots with $\gamma$ far from $0$) the best-fit values of $\Omega_\text{M}^{0}$ are remarkably smaller than in $\Lambda$CDM.}
\end{figure}

{\bf $\bullet$ $\pmb{\beta_1\beta_2}$ and $\pmb{\beta_1\beta_3}$ models}: Looking at the four upper panels of Fig.~\ref{fig:twoparam_plot1} for these models, we observe an interesting feature. The points in the parameter space of both models for which $|(c_g^2-1)(c_f^2-1)|$ is small seem to be residing on a thin region, shown with shades of black. All the other points are excluded by gravitational waves, although they give good fits to the cosmological observations. Let us try to understand this favored, thin region. We argued in the previous section that if $r$ becomes equal to $\gamma$, even at one point over the history (in addition to far in the past), the two background metrics of the model should be proportional at all times. This means that in particular if a point in the parameter space requires $r=\gamma$ at the present time, that point should correspond to proportional metrics. Now looking at the plots of $r$ versus $\gamma$ for both $\beta_1\beta_2$ and $\beta_1\beta_3$ models, we see that the very thin, linelike part of the favored region is indeed the $r=\gamma$ line. This therefore shows that one main region with $(c_g^2-1)(c_f^2-1)\approx0$ corresponds in fact to the cases with proportional backgrounds. This can be seen further by looking at the plots of $b$ versus $\gamma$. The thin, black line now corresponds to $b=1$, as expected for proportional metrics. The other tiny region with $(c_g^2-1)(c_f^2-1)$ being very small is the one in the vicinity of $\gamma=0$. Note that this region is not clearly visible in the plots because it is a highly thin region perpendicular to the $\gamma$ axis and is difficult to depict. The plots are therefore consistent with our analytical arguments in the previous section that only singly coupled submodels or the ones with the two background metrics being proportional are consistent with the speed of gravitational waves being the same as the speed of light. The observations of gravitational waves therefore highly constrain these two bigravity models as it was the case also for the single-parameter models. Note that the upper cuts in the plots are the result of the finite ranges which we have chosen in our MCMC scans for the $\beta_n$ parameters. We have checked that by increasing these ranges the cuts on the plots systematically move upwards, but the main features do not change---the thin, favored regions  only extend to larger  $\gamma$ and $r$. Finally, we show in the upper panels of Fig.~\ref{fig:twoparam_plot2} the constraints on $\Omega_\text{M}^{0}$, the present value of the matter density parameter, for the $\beta_1\beta_2$ and $\beta_1\beta_3$ models. We can clearly see that there are two regions with $(c_g^2-1)(c_f^2-1)$ being close to $0$, one in the vicinity of $\gamma=0$, corresponding to the singly coupled corner of the theory, and the other one with $\gamma$ far from $0$, corresponding to proportional backgrounds. It is interesting to note that the values of $\Omega_\text{M}^{0}$ for the latter case which are consistent with GW constraints are significantly smaller than the best-fit value of $\sim 0.3$ for the $\Lambda$CDM model.

{\bf $\bullet$ $\pmb{\beta_0\beta_1}$ and $\pmb{\beta_0\beta_2}$ models}: Let us now investigate the two $\beta_0\beta_1$ and $\beta_0\beta_2$ models, by studying the four lower panels of Fig.~\ref{fig:twoparam_plot1}. Overall, the same features as in the previous models of $\beta_1\beta_2$ and $\beta_1\beta_3$ can be seen here, especially that proportional backgrounds survive the bounds on the speed of gravitational waves. This can be seen again as a thin $r=\gamma$ line. There is however an interesting difference in these two models compared to the previous ones.

The parameters $\beta_1$ and $\beta_2$ being $0$ in each case while $\gamma$ is also set to $0$ corresponds to $\Lambda$CDM, with $\beta_0$ playing the role of the cosmological constant. We may therefore expect a large concentration of cosmologically viable points in the $\gamma\approx0$ region. Even though this region does exist, as is better visible for the $\beta_0\beta_1$ model, the majority of the viable points seem to be clustering around large $\gamma$, especially for the $\beta_0\beta_2$ model. In order to understand this, let us look at Figs.~\ref{fig:beta0_plot1} and~\ref{fig:beta2_plot1} for the single-parameter, $\beta_0$ and $\beta_2$ models. It is clear from these figures that the models act in opposite ways. While the $\beta_0$ model favors small $\gamma$, the $\beta_2$ model does not admit $\gamma$ smaller than $\sim 0.3$. Although we may expect the entire range of $\gamma$ to be covered by turning on both of the parameters, our numerical investigations show that the points in the parameter space of the $\beta_0\beta_2$ model fit the cosmological observations better when $\beta_0$ is not $0$ and $\gamma$ is large. That is why the density of the points in the figures is higher at large $\gamma$, where the model deviates significantly from the singly coupled scenario. The same holds for the $\beta_0\beta_1$ model, although in that case the singly coupled submodel is less disfavored. This can be understood by looking at Fig.~\ref{fig:beta1_plot1} for the single-parameter, $\beta_1$ model, where the plots show that small $\gamma$ are cosmologically viable, contrary to the $\beta_2$ model.
\begin{figure*}
\center
  \includegraphics[height=3.8cm]{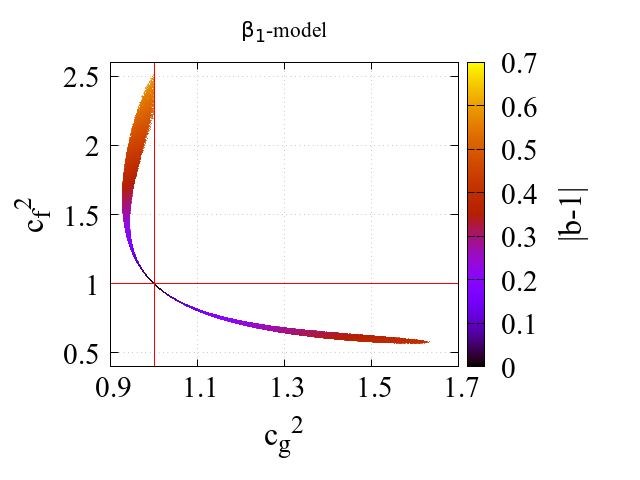}
  \includegraphics[height=3.8cm]{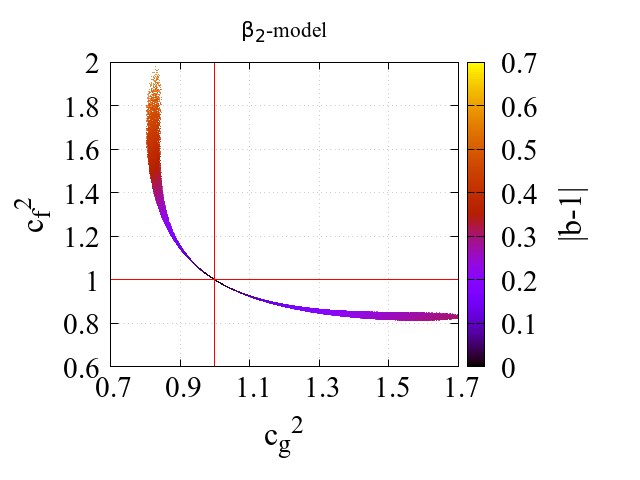}
  \includegraphics[height=3.8cm]{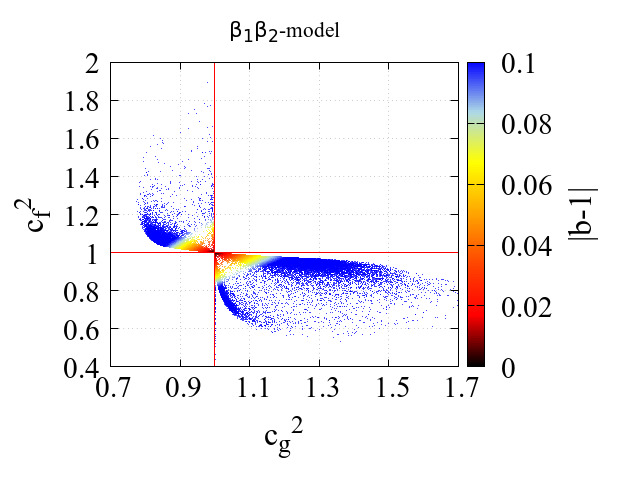}
  \includegraphics[height=3.8cm]{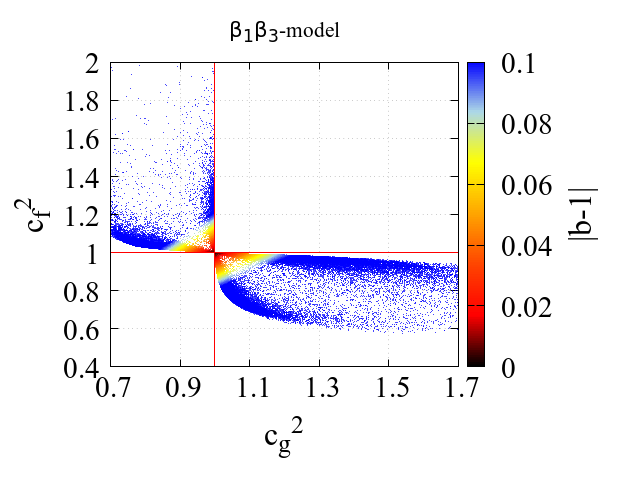}
  \includegraphics[height=3.8cm]{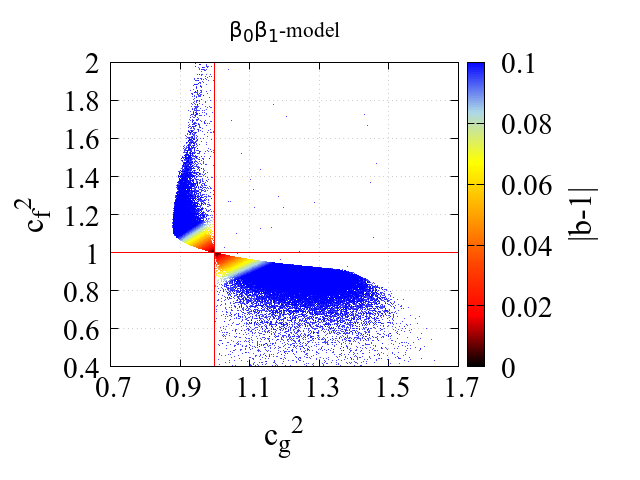}
  \includegraphics[height=3.8cm]{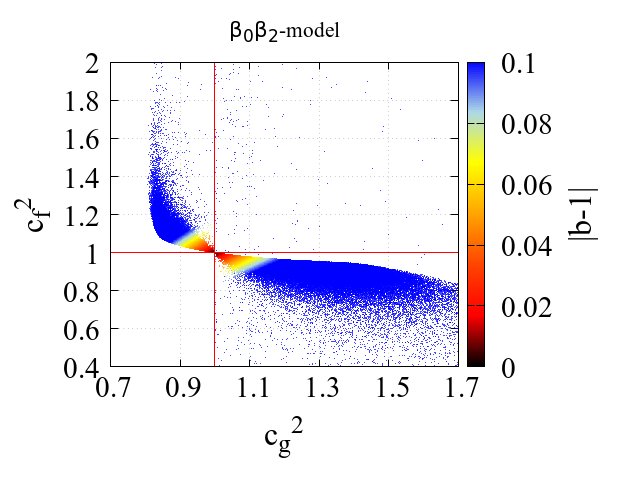}
\caption{\footnotesize\label{fig:twoparam_plot3} Scatter plots showing the values of the speed of gravitational waves for the tensor modes corresponding to the two metrics $g_{\mu\nu}$ and $f_{\mu\nu}$ for the two-parameter models of $\beta_1\beta_2$, $\beta_1\beta_3$, $\beta_0\beta_1$, and $\beta_0\beta_2$, as well as the single-parameter $\beta_1$ and $\beta_2$ models. The color shows the value of $|b-1|$ at each point in the parameter space, as a measure of the deviation from proportional backgrounds (with $b=1$). The red, vertical and horizontal lines show $c_g=1$ and $c_f=1$, respectively. Again, all the quantities have been computed at the present time ($z=0$).}
\end{figure*}

\subsection{Further remarks}

Before we end the discussions of our numerical investigation, let us present the results of our MCMC scans for all the two-parameter models of $\beta_1\beta_2$, $\beta_1\beta_3$, $\beta_0\beta_1$, and $\beta_0\beta_2$, as well as the single-parameter models of $\beta_1$ and $\beta_2$, now in terms of the speed of the gravitational waves corresponding to the two metrics of the theory, $g_{\mu\nu}$ and $f_{\mu\nu}$. These have been shown in Fig.~\ref{fig:twoparam_plot3}. In order to see how far each cosmologically viable point in the parameter space is from the proportional backgrounds, we color code the points by the value of $|b-1|$. All the quantities $c_g$, $c_f$, and $b$ have been computed at the present time, i.e. at $z=0$.

First of all, the plots confirm our analytical arguments in the previous section that having $c_g=1$ ($c_f=1$) automatically implies $c_f=1$ ($c_g=1$), unless the theory is singly coupled. In addition, the plots also show that $c_f=c_g=1$ is equivalent to $b=1$, i.e. it corresponds to proportional backgrounds, as expected. These can clearly be seen in all the panels. Let us first focus on the single-parameter cases of $\beta_1$ and $\beta_2$, i.e. the first two upper panels of Fig.~\ref{fig:twoparam_plot3}. The intersections of the $c_g=1$ and $c_f=1$ lines in both models correspond to the proportional backgrounds, as $b=1$ at those points. In addition, for the $\beta_1$ model we see that there are points for which $c_g^2=1$ while $c_f^2$ takes larger values ($\sim 2.3$). This is fully consistent with our previous discussions that the $\beta_1$ model admits cosmologically viable singly coupled solutions---these are the points with $c_g=1$ and therefore consistent with the GW observations. The $\beta_2$ model, on the other hand, does not allow singly coupled models consistent with cosmological observations, and we therefore do not see any points in the $\beta_2$ panel of Fig.~\ref{fig:twoparam_plot3} with $c_g=1$ and $c_f\ne1$. Note that in our analysis where we work with $\gamma$ instead of $\alpha$ and $\beta$, the singly coupled models are captured only by $g_{\mu\nu}$ being the physical metric, as we fix $\alpha$ to unity and therefore $\gamma=\beta$. That is why we do not see any points with $c_f=1$ and $c_g\ne1$ for the $\beta_1$ model. Let us now focus on the two-parameter models. As we discussed above, the $\beta_0\beta_1$ and $\beta_0\beta_2$ models do not favor singly coupled solutions, and that is why we do not see many points in the corresponding panels of Fig.~\ref{fig:twoparam_plot3} with $c_g=1$ and $c_f\ne1$. Out of the two other two-parameter models of $\beta_1\beta_2$ and $\beta_1\beta_3$, we see that in the latter case there is a concentration of cosmologically favored points along the vertical line of $c_g^2=1$ even with $c_f^2\ne1$ in the $\beta_1\beta_2$ and $\beta_1\beta_3$ panels of Fig.~\ref{fig:twoparam_plot3}. This is again consistent with our findings above that singly coupled bigravity is not disfavored in the $\beta_1\beta_3$ model.

\section{Conclusions}\label{conclusions}

In this paper, we have extensively studied the implications of the recently detected gravitational waves from a neutron star merger and their electromagnetic counterpart on the viability of the doubly coupled theory of bimetric gravity, and have identified the regions of the parameter space that are consistent with both cosmological observations and gravitational wave measurements. We have been interested in models that provide an alternative explanation for the late-time acceleration of the Universe, and therefore require an interaction (or mass) scale of the order of the present value of the Hubble parameter (i.e. $m\sim H_0$). Our studies have been based on both an analytical investigation of cosmic evolution and propagation of tensor modes in the theory, as well as a numerical exploration of the parameter space of the model using MCMC scans. We have demonstrated that the only regions of the parameter space that survive both the cosmological and gravitational wave constraints are those with the two background metrics being proportional or the singly coupled submodels. Our findings therefore demonstrate that the theory is strongly constrained by the bounds on the speed of gravity waves if it is considered as the mechanism behind cosmic acceleration.

The cases with proportional backgrounds are particularly interesting for various reasons~\cite{Schmidt-May:2014xla}. First of all, the background evolution of the Universe as well as linear perturbations mimic those of the $\Lambda$CDM model, and the model is therefore consistent with all the existing cosmological observations. This also means that the model does not suffer from any ghost or gradient instabilities, which are the typical drawbacks of singly coupled cosmological scenarios, in the (visible) sector where the cosmological perturbations are coupled to matter. The model is however expected to deviate from general relativity, and therefore $\Lambda$CDM, at the nonlinear level and in the early Universe such as the radiation era, where a vector instability in the (hidden) sector decoupled from matter would have to be cured by an as yet unknown UV completion. The expected nonlinear deviations from general relativity in the late Universe open up an interesting route for further tests of the theory using the observations of structure formation and evolution at nonlinear scales. In addition, graviton mass eigenstates can be diagonalized only around the proportional backgrounds, and therefore the notion of spin-2 mass makes sense only in those cases---singly coupled bigravity does not admit proportional metrics in the presence of matter. Moreover, the effective metric of the doubly coupled theory, which is the one that couples to matter, corresponds to the massless modes at the linear level, while the massive modes are fully decoupled; the massive and massless modes however mix at the nonlinear level.

We therefore conclude that the recent, tight constraints on the speed of gravitational waves leave us with a highly constrained corner of bigravity which is theoretically healthy at low energies\footnote{These models are valid below the cutoff scale $\Lambda_3$ and are therefore well suited for a description of the late-time Universe.} and observationally viable. It remains to be seen whether the model will also fit the cosmological observations at the nonlinear level, or will be ruled out; we leave the investigation of this interesting question for future work.

\acknowledgments Y.A. acknowledges support from the Netherlands Organization for Scientific Research (NWO) and the Dutch Ministry of Education, Culture and Science (OCW), and also from the D-ITP consortium, a program of the NWO that is funded by the OCW. A.C.D is partially supported by STFC under Grants No. ST/L000385/1 and No. ST/L000636/1. V.V. is supported by a de Sitter PhD fellowship of NWO. This work is supported in part by the EU Horizon 2020 research and innovation program under the Marie-Sklodowska Grant No. 690575. This article is based upon work related to the COST Action Grant No. CA15117 (CANTATA) supported by COST (European Cooperation in Science and Technology).

\appendix

\section{Tensor modes}\label{Tens_modes}

Here we present the detailed derivation of tensor perturbations and their propagation equations in doubly coupled bimetric gravity. We present the calculations in the metric formalism at the level of the equations of motion, as well as at the action level, both in metric and vierbein formalisms.
\vspace{2 mm}

\textbf{\textit{Derivation from equations of motion.}} --- Here our starting point is the full (modified) Einstein equations for the two metrics $g_{\mu\nu}$ and $f_{\mu\nu}$, which are given by (see Ref.~\cite{Schmidt-May:2014xla} for details)
\begin{eqnarray}\label{eq:fieldeq_g}
&\!(\mathbb{X}\!&\!^{-1})^{(\mu}_{\;\;\;\alpha}G_g^{\nu)\alpha}+m^2\sum\limits_{n=0}^{3}(-1)^{n}\beta_{n} g^{\alpha\beta}(\mathbb{X}^{-1})^{(\mu}_{\;\;\;\alpha} Y^{\beta)}_{(n)\nu} =\nonumber \\
&=& \frac{\alpha}{M_{\text{eff}}^2}\sqrt{\frac{\text{det}g_{\text{eff}}}{\text{det}g}}\left(\alpha(\mathbb{X}^{-1})^{(\mu}_{\;\;\;\alpha}T^{\nu)\alpha} + \beta T^{\mu\nu}\right)\,,
\end{eqnarray}
and
\begin{eqnarray}\label{eq:fieldeq_f}
&\!\mathbb{X}\!&\!^{(\mu}_{\;\;\;\alpha}G_f^{\nu )\alpha}+m^2\sum\limits_{n=0}^{3}(-1)^{n}\beta_{4-n} f^{\alpha\beta}\mathbb{X}^{(\mu}_{\;\;\;\alpha} \hat{Y}^{\nu)}_{(n)\beta} =\nonumber \\
&=& \frac{\beta}{M_{\text{eff}}^2}\sqrt{\frac{\text{det}g_{\text{eff}}}{\text{det}f}}\left(\alpha T^{\mu\nu}+\beta\mathbb{X}^{(\mu}_{\;\;\;\alpha}T^{\nu)\alpha}\right)\,,
\end{eqnarray}
where $G_g^{\mu\nu}$ and $G_f^{\mu\nu}$ are the Einstein tensors for $g_{\mu\nu}$ and $f_{\mu\nu}$, respectively, $T^{\mu\nu}$ is the stress-energy tensor corresponding to the effective metric $g^\text{eff}_{\mu\nu}$, and the square-root matrices $\mathbb{X}$ and $\mathbb{X}^{-1}$ are defined through
\begin{eqnarray}
\mathbb{X}^{\mu}_{\;\;\;\alpha}\mathbb{X}^{\alpha}_{\;\;\;\nu} &\equiv &g^{\mu\beta}f_{\beta\nu}\,,\label{eq:X-matrices}\\
(\mathbb{X}^{-1})^{\mu}_{\;\;\;\alpha}(\mathbb{X}^{-1})^{\alpha}_{\;\;\;\nu} &\equiv& f^{\mu\beta}g_{\beta\nu} \,.\label{eq:Xm1-matrices}
\end{eqnarray}

Now, the linear metric perturbations for $g$ and $f$ tensor modes $h_{g+/\times}$ and $h_{f+/\times}$ can be written as
\begin{eqnarray}
\dd s_g^2 = -N_g^2 \dd t^2 &+& a_g^2[(1+h_{g+})\dd x^2+(1-h_{g+})\dd y^2\nonumber\\
&+& \dd z^2 + 2h_{g\times}dxdy]\,,\\
\dd s_f^2 = -N_f^2 \dd t^2 &+& a_f^2[(1+h_{f+})\dd x^2+(1-h_{f+})\dd y^2\nonumber\\
&+& \dd z^2 + 2h_{f\times}\dd x \dd y]\,.
\end{eqnarray}

Plugging these into Eqs.~(\ref{eq:X-matrices}) and (\ref{eq:Xm1-matrices}) we find
\begin{equation}\label{eq:Xtens}
\mathbb{X}^{\alpha}_{\;\;\beta}\!=\!\left(
\begin{array}{cccc}
 \frac{N_f}{N_g}\!&\!0\!&\!0\!&\!0 \\
\!0\!&\!\frac{a_f}{a_g}\!+\!\frac{a_f}{a_g}\!\frac{(h_{f+}\!-\!h_{g+})}{2}\!&\!\frac{a_f}{a_g}\!\frac{(h_{f\times}\!-\!h_{g\times})}{2}\!&\!0 \\
\!0\!&\!\frac{a_f}{a_g}\!\frac{(h_{f\times}\!-\!h_{g\times})}{2}\!&\!\frac{a_f}{a_g}\!+\!\frac{a_f}{a_g}\!\frac{(h_{f+}\!-\!h_{g+})}{2}\!&\!0 \\
\!0\!&\!0\!&\!0\!&\!\frac{a_f}{a_g} \\
\end{array}
\right)\,,
\end{equation}
and
\begin{equation}\label{eq:Xm1tens}
(\mathbb{X}^{-1})^{\alpha}_{\;\;\beta}\!=\!\left(
\begin{array}{cccc}
 \frac{N_g}{N_f}\!&\!0\!&\!0\!&\!0 \\
\!0\!&\!\frac{a_g}{a_f}\!-\!\frac{a_g}{a_f}\!\frac{(h_{f+}\!-\!h_{g+})}{2}\!&\!\frac{a_g}{a_f}\!\frac{(h_{g\times}\!-\!h_{f\times})}{2}\!&\!0 \\
\!0\!&\!\frac{a_g}{a_f}\!\frac{(h_{g\times}\!-\!h_{f\times})}{2}\!&\!\frac{a_g}{a_f}\!-\!\frac{a_g}{a_f}\!\frac{(h_{f+}\!-\!h_{g+})}{2}\!&\!0 \\
\!0\!&\!0\!&\!0\!&\!\frac{a_g}{a_f} \\
\end{array}
\right)\,,
\end{equation}
for the square-root matrices at the linear order.

Having these expressions for $\mathbb{X}$ and $\mathbb{X}^{-1}$, the nonvanishing parts of the tensor sector of the effective metric can be shown to be
\begin{eqnarray}\label{eq:eff_mode_1}
\delta g^{\text{eff}}_{11} &=& -\delta g^{\text{eff}}_{22} \equiv a^2 h_{\text{eff}+} \nonumber\\
&=& a\left(\alpha a_g h_{g+} + \beta a_f h_{f+}\right)\,,\\
\delta g^{\text{eff}}_{12} &=& \delta g^{\text{eff}}_{21} \equiv a^2 h_{\text{eff}\times} \nonumber \\
&=& a\left(\alpha a_g h_{g\times} + \beta a_f h_{f\times}\right)\,.\label{eq:eff_mode_2}
\end{eqnarray}

By using Eqs. (\ref{eq:Xtens}) and (\ref{eq:Xm1tens}) in the field equations we recover Friedmann equations at the background level, while at the linear order we obtain the propagation equations for the tensor modes $h_{g+/\times}$ and $h_{f+/\times}$,
\begin{widetext}
\begin{eqnarray}
\frac{1}{N_g^2}\ddot{h}_{g+/\times} &+& (3\frac{H_g}{N_g}-\frac{\dot{N}_g}{N_g^3})\dot{h}_{g+/\times} - \frac{1}{a_g^2}\nabla^2 h_{g+/\times}+ A(h_{f+/\times} - h_{g+/\times}) = 0\,,\\
\frac{1}{N_f^2}\ddot{h}_{f+/\times} &+& (3\frac{H_f}{N_f}-\frac{\dot{N}_f}{N_f^3})\dot{h}_{f+/\times} - \frac{1}{a_f^2}\nabla^2 h_{f+/\times} + B(h_{g+/\times} - h_{f+/\times}) = 0\,,
\label{eqah}
\end{eqnarray}
where
\begin{eqnarray}
A &\equiv & r\frac{1}{M_\text{eff}^2} \left(\alpha \beta p (\alpha +\beta r) \left(\alpha +\frac{\beta  N_f}{N_g}\right) - m^2 M_\text{eff}^2 \left(\beta_1+\frac{N_f \left(\beta_2+\beta_3 r\right)}{N_g}+\beta_2 r\right)\right)\,,\\
B &\equiv & \frac{1}{r}\frac{1}{M_\text{eff}^2} \left(\alpha \beta p (\beta +\alpha \frac{1}{r}) \left(\beta +\frac{\alpha  N_g}{N_f}\right)- m^2 M_\text{eff}^2 \left(\beta_3+\frac{N_g \left(\beta_2+\beta_1 \frac{1}{r}\right)}{N_f}+\beta_2 \frac{1}{r}\right)\right)\,,
\end{eqnarray}
\end{widetext}
with $p$ here being the pressure of the matter sector.

It should be noted that these two propagation equations can be written in a form that manifestly shows the symmetry of the interaction terms (i.e. the symmetry of the mass matrix). This can be seen by rewriting the propagation equations as
\begin{eqnarray}
\!\frac{d}{dt}\!\left(\!\frac{a_g^3}{N_g}\!\dot{h}_{g+/\times}\!\right)\!&\! - \!&\!a_g^3N_g\!\frac{1}{a_g^2}\nabla^2h_{g+/\times}\nonumber\\
&\!+\!&\!a_g^3N_gA(h_{f+/\times}\! - \!h_{g+/\times})\!=\!0\,,\label{eq:symm_tensor_eqns_1}\\
\frac{d}{dt}\!\left(\!\frac{a_f^3}{N_f}\!\dot{h}_{f+/\times}\!\right)\!&\!-\!&\!a_f^3N_f\!\frac{1}{a_f^2}\nabla^2h_{f+/\times}\nonumber\\
&\!+\!&\!a_g^3N_gA(h_{g+/\times}\! - \!h_{f+/\times})\!=\!0\,,\label{eq:symm_tensor_eqns_2}
\end{eqnarray}
where now the same factor of $a_g^3N_gA$ appears in front of $h_{f+/\times}$ in the first equation and in front of $h_{g+/\times}$ in the second equation.
\vspace{2 mm}

\textbf{\textit{Derivation of the quadratic action.}} --- In order to facilitate the comparison with the results of Refs.~\cite{Brax:2016ssf,Brax:2017hxh} let us also present the calculation of the graviton mass matrix at the level of the action. In this analysis we ignore the matter sector, i.e. we study a fully dark energy dominated epoch.

First of all, by varying the background part of the action with respect to the lapses and scale factors we recover the background equations of motion
\begin{eqnarray}
\label{eq:bckgr_eom_1}
3H_g^2 &=& m^2B_0\,,~~~~~~3H_f^2 = m^2 B_1\,,\\
\label{eq:bckgr_eom_2}
\ddot{a}_g &=& \frac{1}{2} m^2 a_g N_g^2 \left(B_0 + (\beta_1 + 2\beta_2 r + \beta_3 r^2) \left( \frac{N_f}{N_g}-r\right)\right)\nonumber\\
&&+a_g H_g \dot{N}_g-\frac{1}{2} a_g H_g^2 N_g^2\,,\\
\label{eq:bckgr_eom_3}
\ddot{a}_f &=& \frac{1}{2} m^2 a_f N_f^2 \left(B_1 + (\beta_3 + 2\frac{\beta_2}{r} + \frac{\beta_1}{r^2}) \left( \frac{N_g}{N_f}-\frac{1}{r}\right)\right)\nonumber\\
&&+a_f H_f \dot{N}_f-\frac{1}{2} a_f H_f^2 N_f^2\,.
\end{eqnarray}

Our objective here is to obtain the mass terms of the gravitational waves. In principle, the calculation of the quadratic action is straightforward, but the subtle point here is that besides the potential terms of bigravity, also the two Einstein-Hilbert terms contribute with additional terms quadratic in $h_{g+/\times}$ and $h_{f+/\times}$. Let us exemplify this by looking at the kinetic term of the $g$-sector. First of all, there is a contribution from the volume factor, which reads as
\begin{equation}
S^{(2)} \supset -\frac{M^2_{\text{eff}}}{2}\int{\dd^4x \left( -\frac{N_g a_g^3}{2} (h_{g\times}^2+h_{g+}^2)\right) \bar{R}_g}\,,
\end{equation}
where $\bar{R}_g$ is the background part of the Ricci scalar, which is given by
\begin{equation}
\bar{R}_g = 6\frac{a_g N_g \ddot{a}_g-a_g \dot{a}_g\dot{N}_g+N_g \dot{N}_g^2}{a_g^2 N_g^3}\,.
\end{equation}
Additional contributions come from some of the terms in the perturbed part of the Ricci scalar, namely from
\begin{eqnarray}
S^{(2)} \supset -\frac{M^2_{\text{eff}}}{2}\int\dd^4x [ f(t)(h_{g+}\dot{h}_{g+}&+&h_{g\times}\dot{h}_{g\times})+\nonumber\\
F(t)(h_{g+}\ddot{h}_{g+}&+&h_{g\times}\ddot{h}_{g\times})]\,,
\end{eqnarray}
where
\begin{equation}\label{eq:fF}
f(t)=\frac{a_g}{N_g^2}\left( 2 a_g^2 \dot{N}_g-8 a_g N_g \dot{a}_g \right),\,F(t)=-2\frac{a_g^3}{N_g}\,.
\end{equation}
The corresponding contributions to the mass matrix are given by
\begin{equation}
S^{(2)} \supset -\frac{M^2_{\text{eff}}}{2}\int\dd^4x \frac{\ddot{F}(t)-\dot{f}(t)}{2}(h_{g+}^2+h_{g\times}^2)\,.
\end{equation}
Note that we needed to divide by a factor of $2$ in the last expression, because in the original terms only the variations with respect to the fields under the time derivatives could contribute to the mass terms in the equations of motion.

These contributions should be added to the contributions from the potential terms. In order to find the latter we also need the second-order piece of the $\mathbb{X}^{\mu}_{\;\;\;\nu}$ matrix, the nonvanishing components of which are found to be
\begin{eqnarray}
\delta^{(2)}\mathbb{X}^{1}_{\;\;1} &=& \delta^{(2)}\mathbb{X}^{2}_{\;\;2}=\nonumber\\
 &=& -r \sum_{\star=\times,+}\frac{h^2_{f\star} - 3h^2_{g\star} + 2h_{f\star}h_{g\star}}{8}\,,\\
\delta^{(2)}\mathbb{X}^{1}_{\;\;2} &=& \delta^{(2)}\mathbb{X}^{2}_{\;\;1} = -r\frac{h_{f\times} h_{g+} - h_{g\times} h_{f+}}{2}\,.
\end{eqnarray}

Combining all the potential terms and dropping an overall factor of $1/2$ from the action we obtain the graviton mass terms
\begin{equation}\label{eq:mass_action_metric}
S^{(2)} \supset M^2_{\text{eff}}\int\dd^4x \frac{1}{2}\sum_{\star=\times,+} \mathbb{M}^{IJ} h_{I\star}h_{J\star}\,,
\end{equation}
where the mass matrix is found to be
\begin{eqnarray}\label{eq:mmaa}
\mathbb{M}^{gg}&=& \mathbb{M}^{ff} = -\mathbb{M}^{gf} = -\mathbb{M}^{fg}=\nonumber\\
 &=& m^2 a_g^3 N_g r\!\left(\!\beta_1\!+\!\beta_2(\frac{N_f}{N_g}\!+\!r)\!+\!\beta_3 \frac{N_f}{N_g} r\!\right).
\end{eqnarray}
Note particularly that we have recovered the same interaction terms as in Eqs.~(\ref{eq:symm_tensor_eqns_1}) and (\ref{eq:symm_tensor_eqns_2}).

In Refs.~\cite{Brax:2016ssf,Brax:2017hxh} the interaction sector has been written in terms of the constrained metric vierbeins as
\begin{eqnarray}
S_\text{interaction} &=& m^2 M^2_{\text{eff}}\!\sum_{IJKL}m^{IJKL}\times\nonumber\\
&&\!\times\!\int\!d^4x \epsilon_{abcd}\epsilon^{\mu\nu\rho\sigma}e_{I\mu}^ae_{J\nu}^be_{K\rho}^ce_{L\sigma}^d,
\label{poo}
\end{eqnarray}
where the tetrad fields (or vierbeins) are defined through
\begin{equation}\label{eq:vierbeins}
g_{\mu\nu}^I = \eta_{ab}e_{I\mu}^ae_{I\nu}^b\,.
\end{equation}
Here $I$ labels the two metrics, $I = \{g,f\}$, $\mu$ and $\nu$ are the covariant indices, and $a$ and $b$ are the indices in the local Lorentz frame. The interaction matrix $m^{IJKL}$ is fully symmetric and its components in terms of the $\beta_{0,...,4}$ parameters are given by
\begin{eqnarray}
m^{gggg} &=& \frac{\beta_0}{24}\,,~~m^{fggg} = \frac{\beta_1}{24}\,,\\
m^{ffgg} = \frac{\beta_2}{24}\,,~~m^{fffg} &=& \frac{\beta_3}{24}\,,~~m^{ffff} = \frac{\beta_4}{24}\,,
\end{eqnarray}
with the other components being trivially related to the ones above due to the total symmetry of the $m^{IJKL}$ matrix.

In order to derive the mass sector of the quadratic action in the vierbein formalism we first derive the tensor perturbations of the vierbeins by linearizing Eq.~(\ref{eq:vierbeins}). As a result, for the $e_{I\mu}^a$ matrix we have

\begin{equation}
e_I\!=\!\left(
\begin{array}{cccc}
 N_I\!&\!0\!&\!0\!&\!0 \\
\!0\!&\!a_I(1 + \frac{1}{2}h_{I+})\!&\!\frac{a_I}{2}h_{I\times}\!&\!0 \\
\!0\!&\!\frac{a_I}{2}h_{I\times}\!&\!a_I(1 -\frac{1}{2}h_{I+})\!&\!0 \\
 0\!&\!0\!&\!0\!&\!a_I \\
\end{array}
\right)\,.
\end{equation}

The total mass matrix is built up from two different parts of the action as before.

The first (diagonal) contribution comes from the Einstein-Hilbert terms in the action, and is given by
\begin{equation}
S_{\text{masses,\,EH}}^{(2)} = -\frac{M^2_{\text{eff}}}{2}\int\dd^4x \sum_{{\star}=\times,+} \delta m^2_{gg} h^g_{\star} h^g_{\star} + (g \rightarrow f)\,,
\end{equation}
where we have found that
\begin{eqnarray}
\delta m^2_{gg} &=& -\frac{N_g a_g^3}{4} \bar{R}_g - \frac{\ddot{F}(t)-\dot{f}(t)}{4}\,,\\
\delta m^2_{ff} &=& \delta m^2_{gg}(g \rightarrow f)\,.
\end{eqnarray}
Here $F(t)$ and $f(t)$ are the same functions as in Eq.~(\ref{eq:fF}).

The second part comes from the expansion of the potential term (\ref{poo}) to second order in the gravitons. Direct calculation gives
\begin{equation}
S_{\text{masses,\,pot}}^{(2)} = \frac{1}{4} m^2 M^2_{\text{eff}}\int\dd^4x \sum_{\star = \times,+} \hat m^2_{IJ} h^I_{\star} h^J_{\star}\,,
\end{equation}
where
\begin{eqnarray}
\hat m^2_{gg} &=& N_g a_g^3\left( \beta_2 r \frac{N_f}{N_g} + \beta_1 r + \beta_1 \frac{N_f}{N_g} + \beta_0 \right),\\
\hat m^2_{ff} &=& N_f a_f^3\left( \beta_3 \frac{1}{r} + \beta_2 \frac{1}{r} \frac{N_g}{N_f} + \beta_4 + \beta_3 \frac{N_g}{N_f} \right),\\
\hat m^2_{fg} &=& \hat m^2_{gf}\nonumber\\
 &=& N_g a_g^3 r \left( \beta_2 \frac{N_f}{N_g} + \beta_3 r \frac{N_f}{N_g} + \beta_1 + \beta_2 r \right).
\end{eqnarray}
Adding the two sectors, making use of the background equations of motion (\ref{eq:bckgr_eom_1})-(\ref{eq:bckgr_eom_3}), and dropping an overall factor of $1/2$ from the action, we retrieve the action (\ref{eq:mass_action_metric}) with the mass matrix given exactly by (\ref{eq:mmaa}).

\textbf{\textit{The massless and massive modes.}} --- The dynamics of the two gravitons can be better understood by switching to the canonically normalized basis
\begin{equation}
h_{I\star}= D_I \bar h_{I\star}\,,
\end{equation}
where $\star= +/\times$ and we have defined
\begin{equation}
D_I \equiv \left( \frac{N_I}{a_I^3} \right)^{1/2}\,.
\end{equation}
In this new basis the mass matrix reads
\begin{equation}
\bar{\mathbb{M}}={\cal M}^2 \left(
\begin{array}{cc}
 D_g^2& -D_g D_f \\
-D_g D_f& D_f^2\\
\end{array}
\right)\,,
\end{equation}
where ${\cal M}^2= \mathbb{M}^{gg}$.
In this basis the graviton equations read
\begin{equation}\label{eq:normalized_eoms}
\ddot{\bar{h}}_{I\star} - c^2_I \frac{N^2}{a^2} \nabla^2 \bar{h}_{I\star} +\bar{\mathbb{M}}^{IJ}\bar{h}_{J\star} - D_I\frac{\dd^2}{\dd t^2}\left( \frac{1}{D_I}\right) \bar{h}_{I\star}=0\,,
\end{equation}
where we have identified the speeds of the waves in the effective conformal time (for which photons have a normalized speed $c_\gamma=1$),
\begin{equation}
c_I= \frac{aN_I}{a_I N}.
\end{equation}

It is easy to see that this mass matrix always has a massless and a massive eigenmode given by
\begin{equation}
\bar V_0= \left(
\begin{array} {c}
1\\
D_g/D_f\\
\end{array}
\right)\,,\,\,
\bar V_m= \left(
\begin{array} {c}
1\\
-D_f/D_g\\
\end{array}
\right)\,,
\end{equation}
with eigenmass square being
\begin{equation}
M^2a^2 = {\cal M}^2 (D_g^2 + D_f^2)\,,
\end{equation}
where the factor of $a^2$ has been included to comply with the usual definition for the mass of graviton in FLRW spacetimes.
In the case of proportional metrics, when $r = \gamma$, the above mass eigenvectors reduce to
\begin{equation}
\bar{V}_0 = \left(
\begin{array} {c}
1\\
\gamma\\
\end{array}
\right),\,\,
\bar{V}_m= \left(
\begin{array} {c}
1\\
-\gamma^{-1}\\
\end{array}
\right)\,,
\end{equation}
which guarantees that one can diagonalize the system of dynamical equations (\ref{eq:normalized_eoms}) by simply adding linear combinations of the two propagation equations with constant coefficients.

Now, one can see that the canonically normalized massless eigenmode is associated to the effective graviton modes. Indeed, first of all from Eqs.~(\ref{eq:eff_mode_1}) and (\ref{eq:eff_mode_2}) we see that $h_{\rm eff} = \alpha D(\bar{h}_g + \gamma \bar{h}_f)$, with $D \equiv \sqrt{N/a^3}$. The canonically normalized version of this field is the massless mode $\bar{h}_0 \equiv \bar{h}_g + \gamma \bar{h}_f$. The massive mode, on the other hand, corresponds to the difference $\bar{h}_m = \bar{h}_g - \bar{h}_f/\gamma$.

Combining the equations of motion in~(\ref{eq:normalized_eoms}) appropriately, we obtain
\begin{eqnarray}
\ddot{\bar{h}}_{0\star} &-& \nabla^2 \bar{h}_{m\star} - \frac{\ddot{a}}{a} \bar{h}_{0\star}=0\,,\\
\ddot{\bar{h}}_{m\star} &-&  \nabla^2 \bar h_{m\star} + ( M^2 a^2 - \frac{\ddot a}{a}) \bar h_{m\star}=0\,.
\end{eqnarray}
Here we have used the fact that for the proportional backgrounds we have $D_I=a_I^{-1}$ if we pick the lapses as $N_I = a_I$. Moreover, recalling that
\begin{equation}\label{bac}
a_g=\frac{\alpha}{\alpha^2 +\beta^2}a\,,\ \ a_f=\frac{\beta}{\alpha^2 +\beta^2}a\,,
\end{equation}
we see that $D_I \dd^2\left( D_I^{-1} \right)/\dd t^2 = \ddot{a}/a$.
The first of these dynamical equations is the propagation equation of gravitons in general relativity, with the gravitons being massless but receiving a ``pseudo''mass of the form $-\ddot a/a$. The second one is the propagation equation for a massive graviton of mass $M$. Notice that for both modes the speed of propagation is $1$, and that (\ref{bac}) implies that the light cones for gravitons and photons
coincide. 

\bibliography{bibliography}

\end{document}